\documentclass[12pt,a4paper]{article}
\pdfoutput=1

\usepackage{jheppub}
\usepackage{dsfont}
\usepackage{amssymb,amsmath}
\usepackage{epsfig}
\usepackage{epstopdf}
\usepackage{latexsym}
\usepackage{graphicx}
\usepackage{subfig}
\usepackage{booktabs}
\usepackage{bbm}
\usepackage{color}

\makeatletter
\def\@fpheader{\relax}
\makeatother

\def\be{\begin{equation}}
\def\ee{\end{equation}}

 \newcommand{\bea}{\begin{eqnarray}}
\newcommand{\eea}{\end{eqnarray}}

 \def\ep{{\epsilon}}

 \def\frac#1#2{{#1\over #2}}

 \def\s{\sqrt}

 \def\Re{{\rm Re}}
 
\def\be{\begin{equation}}
\def\ee{\end{equation}}
\def\ba{\begin{eqnarray}}
\def\ea{\end{eqnarray}}
 \def\ep{\epsilon}

 \def\de{\partial}

 \def\f {\frac}
 \def\ti{\tilde}
 \def\ap{\alpha}

 \def\ddd{\cdot\cdot\cdot}
 \def\no{\nonumber \\}

 \def\la{\langle}
 \def\lb{\rangle}
 \def\ep{\epsilon}

 \def\vp{\varphi}

\usepackage{color}
\usepackage{graphicx}
\usepackage{dcolumn}
\usepackage{bm}

\preprint{\begin{flushright}
YITP-17-65 \\
IPMU17-0091
\end{flushright}}

\title{\Large Liouville Action as Path-Integral Complexity:\\ From Continuous Tensor Networks to AdS/CFT}
\author{Pawel Caputa$^{a}$, Nilay Kundu$^{a}$, Masamichi Miyaji$^{a}$,
Tadashi Takayanagi$^{a,b}$ and Kento Watanabe$^{a}$
}
\affiliation{$^{a}$Center for Gravitational Physics, \\
Yukawa Institute for Theoretical Physics (YITP), Kyoto University, \\
Kitashirakawa Oiwakecho, Sakyo-ku, Kyoto 606-8502, Japan.}
\affiliation{$^{b}$Kavli Institute for the Physics and Mathematics of the Universe,\\
University of Tokyo, Kashiwano-ha, Kashiwa, Chiba 277-8582, Japan.}

\abstract{We propose an optimization procedure for Euclidean path-integrals that evaluate CFT wave functionals in arbitrary dimensions. The optimization is performed by minimizing certain functional, which can be interpreted as a measure of computational complexity, with respect to background metrics for the path-integrals. In two dimensional CFTs, this functional is given by the Liouville action. We also formulate the optimization for higher dimensional CFTs and, in various examples, find that the optimized hyperbolic metrics coincide with the time slices of expected gravity duals. Moreover, if we optimize a reduced density matrix, the geometry becomes two copies of the entanglement wedge and reproduces the holographic entanglement entropy.  Our approach resembles a continuous tensor network renormalization and provides a concrete realization of the proposed interpretation of AdS/CFT as tensor networks. The present paper is an extended version of our earlier report arXiv:1703.00456 and includes many new results such as evaluations of complexity functionals, energy stress tensor, higher dimensional extensions and time evolutions of thermofield double states.}

\begin{document}

\maketitle
\flushbottom

\section{Introduction}

The AdS/CFT correspondence \cite{Ma} has been the most powerful tool to understand quantum nature of gravity. Nevertheless, we still do not understand its basic mechanism nor how spacetime in gravity emerges from conformal field theories (CFTs). Recently, possible candidates which might explain the basic mechanism of the AdS/CFT correspondence have begun to be actively investigated. Among them, a very attractive candidate is the idea of emergent spacetimes from tensor networks, as first conjectured by Swingle \cite{Swingle}, for the description of CFT states in terms of MERA (multi-scale entanglement renormalization ansatz) \cite{MERA}\footnote{For recent developments we would like to ask readers to refer to e.g. \cite{cMERA,Beny,NRT,MT,HAPPY,Cz,MNSTW,HQ,MTW}.}. One strong evidence for this correspondence between holography and tensor networks, apart from the symmetry considerations, is the fact that the holographic entanglement entropy formula \cite{RT,HRT} can naturally be explained in this approach by counting the number of entangling links in the networks.

However, up to now, most arguments in these directions have been limited to studies of discretized lattice models so that we can apply the idea of tensor networks directly. Therefore, they at most serve as toy models of AdS/CFT as they do not describe the genuine CFTs which are dual to the AdS gravity (though they provide us with deep insights of holographic principle such as quantum error corrections \cite{HAPPY,ADH}). Clearly, it is then very important to develop a continuous analogue of tensor networks related to AdS/CFT. There already exists a formulation called cMERA (continuous MERA) \cite{cMERA}, whose connection to AdS/CFT has been explored in \cite{NRT,MT,MRTW,MNSTW,MTW}. Nevertheless, explicit
formulations of cMERA are so far only available for free field theories \cite{cMERA}
(see \cite{NRT,HuVidal,FV} for various studies) which is the opposite regime from the strongly interacting CFTs which possess gravity duals, the so-called holographic CFTs. A formal construction of cMERA for general CFTs can be found in \cite{MRTW,MTW}.

The main aim of this work is to introduce and explore a new approach which realizes a continuous limit of tensor networks and allows for field theoretic computations. In our preceding letter
version \cite{CKMTW}, we gave a short summary of our idea and its application to two dimensional (2D) CFTs. Essentially, we reformulate the conjectured relation between tensor networks and AdS/CFT from the viewpoint of Euclidean path-integrals. Indeed, the method called tensor network renormalization (TNR) \cite{TNR} shows that an Euclidean path-integral computation of a ground state wave function can be regarded as a tensor network description of MERA. In this argument, one first discretizes the path-integral into a lattice version and rewrites it as a tensor network. Then, an optimization by contracting tensors and removing unnecessary lattice sites, finally yields the MERA network. The `optimization' here refers to some efficient numerical algorithm.

In our approach we will reformulate this idea, but in such a way that we remain working with the Euclidean path-integral. More precisely, we perform the optimization by changing the structure (or geometry) of lattice regularization. The first attempt in this direction was made in \cite{MTW} by introducing a position dependent UV cut off. In this work, we present a systematic formulation of optimization by introducing a metric on which we perform the path-integral. The scaling down of this metric corresponds to the optimization assuming that there is a lattice site on a unit area cell.

To evaluate the amount of optimization we made, we consider a functional $I_{\Psi}$ of the metric for each quantum state $|\Psi\lb$. This functional,
which might appropriately be called ``Path-integral Complexity'', describes the size of our path-integration and corresponds to the computational complexity in the equivalent tensor network description\footnote{The relevance of computational complexity in holography was recently pointed out and holographic complexity was conjectured to be the volume of maximal time slice in gravity duals \cite{SUR,Susskind} (for recent progresses see e.g. \cite{Alis,Barb,Barbb,CMR,Flory}) and the gravity action in Wheeler-De Witt patch in \cite{BrownSusskind1,BrownSusskind2}
(for recent progresses see e.g.\cite{BrS,Lehner:2016vdi,Fis,Chapman:2016hwi,Reynolds:2016rvl,BrSS,Kim:2017lrw,CaSa,Pa}). We would also like to mention that for CFTs, the behavior of the complexity is very similar to the quantum information metric under marginal deformations as pointed out in \cite{InfoM} (refer to \cite{Alis,Miyaji,Flory} for recent developments), where the metric is argued to be well approximated by the volume of maximal time slice in AdS.}. In 2D CFTs, we can identify this functional $I_{\Psi}$ with the Liouville action. The optimization procedure is then completed by minimizing this complexity functional $I_{\Psi}$, and we argue that the minimum value of $I_{\Psi}$ is a candidate for complexity of a quantum state in CFTs. Below, we will perform a systematic analysis of our complexity functional for various states in 2D CFTs, lower dimensional example of NAdS$_2/$CFT$_1$ (SYK) as well as in higher dimensions, where we will find an interesting connection to the gravity action proposal \cite{BrownSusskind1,BrownSusskind2}.\\
Our new path-integral approach has a number of advantages. Firstly, we can directly deal with any CFTs, including holographic ones, as opposed to tensor network approaches which rely on lattice models of quantum spins. Secondly, in the tensor network description there is a subtle issue that the MERA network can also be interpreted as a de Sitter space \cite{Beny,Cz}, while the refined tensor networks given in \cite{HAPPY,HQ} are argued to describe Euclidean hyperbolic spaces. In our Euclidean approach we can avoid this issue and explicitly verify that the emergent space coincides with a hyperbolic space, i.e. the time slice of AdS.\\
This paper is organized as follows: In section \ref{Formulation}, we present our formulation of an optimization of Euclidean path-integrals in CFTs and relate to the analysis of computational complexity and tensor network renormalization. We will also start with an explicit example for a vacuum of a 2D CFT. In section \ref{Optimization2D}, we will investigate the optimization procedure in 2D CFTs for more general states such as finite temperature states and primary states. In section \ref{HEE2DCFT}, we apply our optimization procedure to reduced density matrices. We show that the holographic entanglement entropy and  entanglement wedge naturally arise from this computation. In section \ref{EMTensor}, we will study the energy stress tensor of our 2D CFTs in the optimization analysis. In section \ref{actionlii}, we explicitly evaluate the Liouville action for the optimized solutions and point out that, due to the conformal anomaly, we need to consider a difference of Liouville action, which corresponds to a relative complexity. In section \ref{AdS2CFT1}, we apply our optimization to one dimensional nearly conformal quantum mechanics like SYK models. In section \ref{HIGHERDIM}, propose and provide various support for generalization of our optimization to higher dimensional CFTs. We also compare our results with existing literature of holographic complexity. In section \ref{DiscussionTFD}, we discuss the time evolution of thermo-field dynamics in 2D CFTs as an example of time-dependent states.  Finally, in section \ref{Concl} we summarize our findings and conclude. In appendix \ref{ap:3dgravity}, we comment on the connection of our approach to an earlier work on the relation between the Liouville theory and 3D gravity. In appendix \ref{holcomlit}, we give a brief summary of the results on holographic complexity in literature, focusing on CFT vacuum states. In appendix \ref{SZcase}, we study the properties of complexity functional in the presence of higher derivatives and in appendix \ref{EELiou}, we discuss connections between entanglement entropy and Liouville field.

\section{Formulation of the Path-Integral Optimization}\label{Formulation}
Here we introduce our idea of optimization of Euclidean path-integrals, which was first presented in our short letter \cite{CKMTW}. We consider a discretized version of Euclidean path-integral which produces a quantum wave functional in QFTs, having in mind a numerical computation of path-integrals. The UV cut off (lattice constant) is written as $\ep$ throughout this paper. The optimization here means the most efficient procedure to perform the path-integral in its discretized form\footnote{Please distinguish our optimization from other totally
different procedures such as the optimization of parameter of tensors in tensor networks. Instead, as we will see later in this section, our optimization changes the tensor network structures as in tensor network renormalization \cite{TNR}.}.  In other words, it is the most efficient algorithm to numerically perform the path-integrals which leads to the correct wave functional.

\subsection{General Formulation}
\begin{figure}[h!]
  \centering
  \includegraphics[width=8cm]{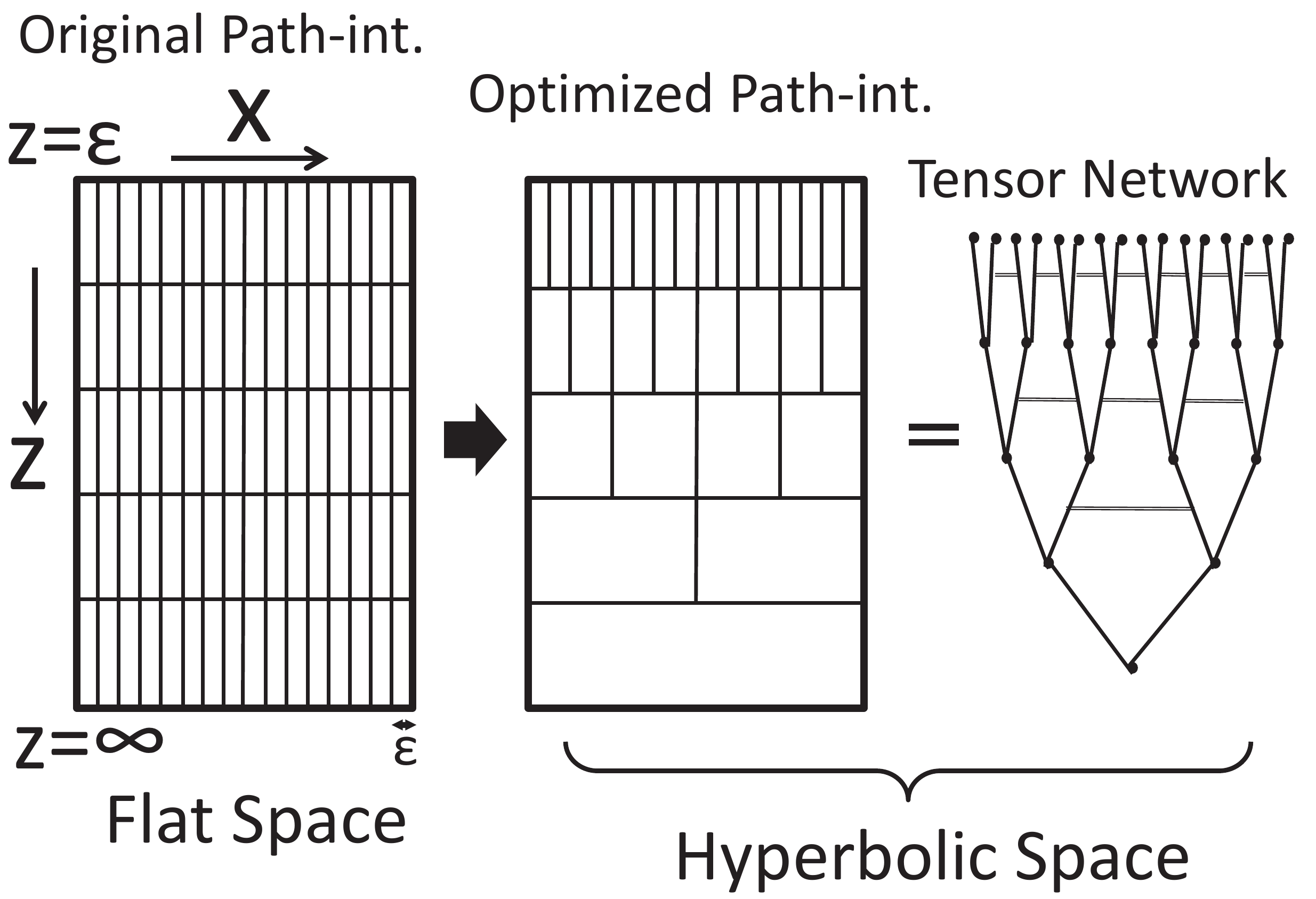}
  \caption{Computation of a ground state wave function from Euclidean path-integral (left) and its optimization (middle), which is described by a hyperbolic geometry. The right figure schematically shows its tensor network expression.}
\label{pathfig}
  \end{figure}

We can express the ground state wave functional in a
$d$ dimensional QFT on $R^d$ in terms of a Euclidean path-integral as follows:
\be
\Psi_0[\ti{\vp}(x)]=\int \left(\prod_{x}\prod_{\ep\leq z< \infty}D\vp(z,x)\right)e^{-S_{QFT}(\vp)}  \times \prod_{x}\delta(\vp(\ep,x)-\ti{\vp}(x)). \label{wfgr}
\ee
Here we write the coordinate of R$^d$ as $(z,x)$, where $-z(\equiv \tau)$ is the Euclidean time and $x$ is the $d-1$ dimensional space coordinate of R$^{d-1}$. We set $z=\ep$ at the final time when the path-integral is completed for our convenience. However, we can shift this value as we like without changing our results as is clear from the time translational invariance.
Now we perform our discretization of path-integral in terms of the lattice constant $\ep$.
We start with the square lattice discretization as depicted in the left picture of Fig.\ref{pathfig}.
To optimize the path-integral we can omit any unnecessary lattice sites from our evaluation. Since only the low energy mode $k\ll 1/\tau$ survives after the path-integral for the period $\tau$, we can estimate that we can combine $O(\tau/\ep)$ lattice sites into one site
without losing so much accuracy. It is then clear that the optimization via this coarse-graining procedure leads to the middle picture in Fig.\ref{pathfig}, which coincides with the hyperbolic plane.

One useful way to systematically quantify such coarse-graining procedures is to introduce a metric on the $d$ dimensional space $(z,x)$ (on which the path integration is performed) such that we arrange one lattice site for a unit area. In this rule, we can write the original flat space metric before the optimization as follows:
\be
ds^2=\f{1}{\ep^2}\cdot\left(dz^2+\sum_{i=1}^{d-1}dx^idx^i\right).   \label{flat}
\ee
Consider now the optimization procedure in this metric formulation. The basic rule is to require that
the optimized wave functional $\Psi_{\mbox{opt}}$
 is proportional to the correct ground state wave function (i.e. the one (\ref{wfgr}) for the metric (\ref{flat}) ) even after the optimization i.e. $\Psi_{\mbox{opt}}[\vp(x)]
 \propto\Psi_{0}[\vp(x)]$. The optimization can then be realized by modifying the background metric for the path-integration
\be \label{metand}
\begin{split}
 ds^2=&g_{zz}(z,x)dz^2+g_{ij}(z,x)dx^i dx^j+2g_{zj}(z,x)dzdx^j,  \\
 g_{zz}(z=&\ep,x)=\ep^{-2}, \ \ g_{ij}(z=\ep,x)=\delta_{ij}\cdot \ep^{-2}, \ \ g_{iz}(z=\ep,x)=0,
\end{split}
\ee
where the last constraints argue that the UV regularization agrees with the original one (\ref{flat}) at the end of the path-integration (as we need to reproduce the correct wave functional after the optimization).

 In conformal field theories, because there are no coupling RG flows, we should be able to complete the optimization only changing the background metric as in (\ref{metand}). However, in non-conformal field theories, actually we need to modify external fields $J$ (such as mass parameter or other couplings of various interactions) in a position dependent way $J(z,x)$. The same is true for CFT states in the presence of external fields.

To finalize the optimization procedure, we should provide a sufficient condition for the metric to be "maximally" optimized. Thus, we assume that for each quantum state $|\Psi\lb$, there exists a functional $I_{\Psi}[g_{ab}(z,x)]$ whose minimization with respect to the metric $g_{ab}$ gives such maximal optimization\footnote{In non-conformal field theories or in the presence of external fields in CFTs, this functional depends on gauge fields
for global currents and scalar fields etc. as $I\left[g_{ab}(z,x),A_{a}(z,x),J(z,x),...\right]$.}. In this way, once we know the functional $I_{\Psi}$, we can finalize our optimization procedure.
 As we will see shortly, in 2D CFTs we can explicit identify this functional $I_{\Psi}[g_{ab}(z,x)]$.
 
\subsection{Connection to Computational Complexity}

At an intuitive level, the optimization corresponds to minimizing the number of path-integral operations in the discretized description. As we will explain in subsection \ref{tnre}, we can
map this discretized Euclidean path-integration into a tensor network computation. Tensor networks are a graphical description of wave functionals in quantum many-body systems in terms of networks of quantum entanglement (see e.g.\cite{CiVe,TG}). The optimization of tensor network was introduced in \cite{TNR}, called tensor network renormalization. We are now considering a path-integral counterpart of the same optimization here. In the tensor network description, the optimization corresponds to minimizing the number of tensors. We can naturally identify this minimized number as a computational complexity of the quantum state we are looking at.

\begin{figure}[h!]
  \centering
  \includegraphics[width=4.5cm]{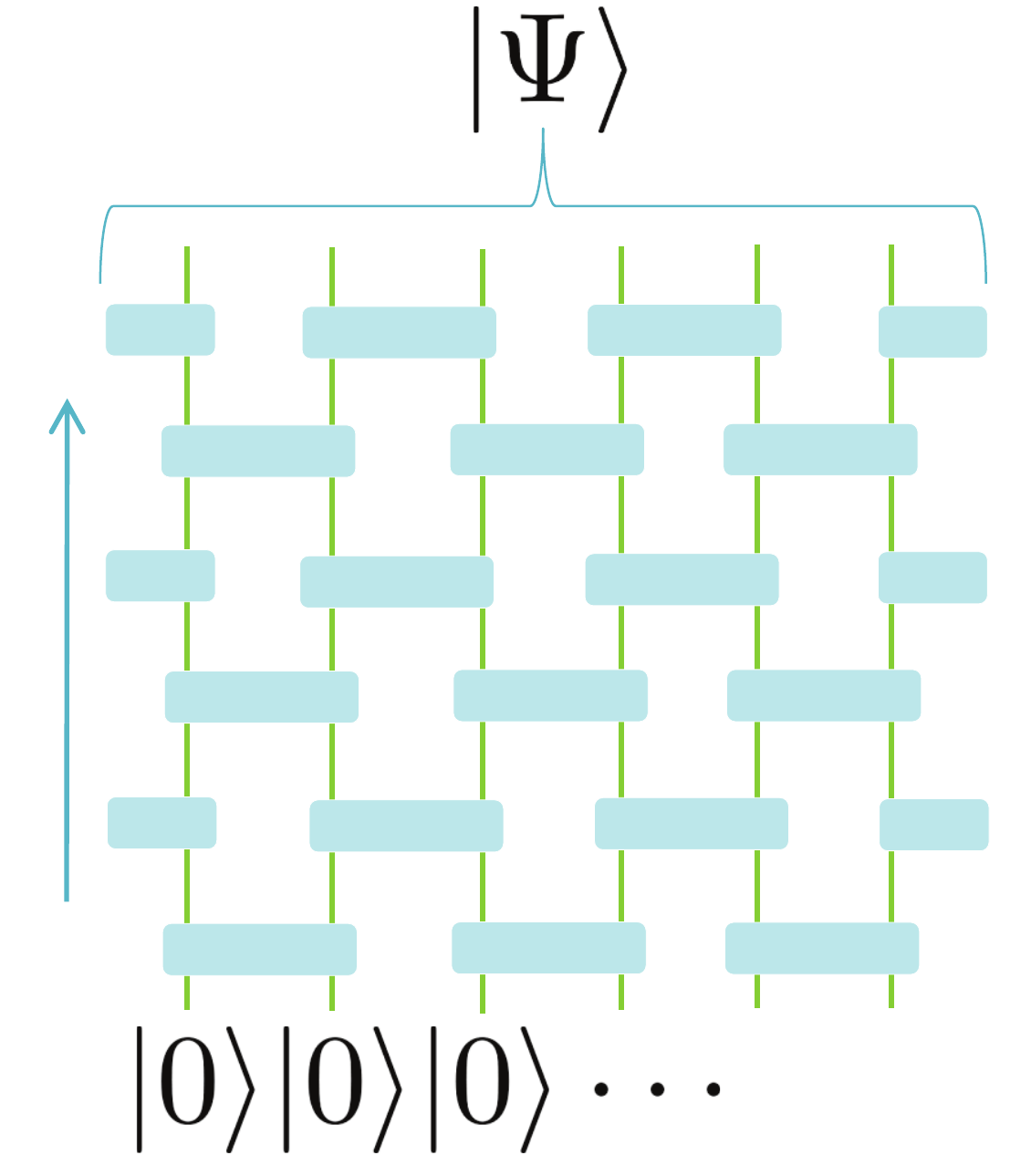}
  \caption{A quantum circuit representation for a quantum state in a qubit system.
A quantum state $| \Psi \rangle$ can be constructed by simple local (2-qubit) unitary operations from a simple reference state, for example, a product state $|0\rangle |0\rangle |0\rangle \cdots$.}
\label{fig:qcircuit}
  \end{figure}

Let us briefly review the relevant facts about the computational complexity of a quantum state (for example, see \cite{CompW, CompO, CompGHLS, CompA}).
In quantum information theory, a quantum state made of qubits can be constructed by a sequence of simple unitary operations acting on a simple reference state.
The sequence is called a quantum circuit and the unitary operations are called quantum gates. As a simple choice, we use 2-qubit gates for simple unitary operations
and a direct product state for a simple state which has no real space entanglement (Fig.\ref{fig:qcircuit}).
The quantum circuit (gate) complexity of a quantum state is then defined as a minimal number of the quantum gates needed to create the state starting from a reference state.
Because the quantum circuit is a model of quantum computation, here we refer to the complexity as the computational complexity \footnote{The relevance of computational complexity in AdS/CFT was recently pointed out and holographic computations of complexity have been proposed in \cite{SUR,Susskind,BrownSusskind1,BrownSusskind2}}.

Based on the above considerations as well as the evidence provided in the following section, we are naturally lead to a conjecture that a computational complexity $C_{\Psi}$ of a state $|\Psi\lb$ is obtained from the functional introduced before by a minimization:
\be
C_{\Psi}=\mbox{Min}_{g_{ab}(z,x)}\left[I_{\Psi}[g_{ab}(z,x)]\right]. \label{cpxidef}
\ee
In other words, the functional $I_{\Psi}[g_{ab}(z,x)]$ for any $g_{ab}(z,x)$ estimates the amount of complexity for that network corresponding to the (partially optimized) path-integral on the space with the specified metric. Understanding of the properties of this complexity functional $I_{\Psi}$, which might appropriately be called ``Path-integral Complexity'',
is the central aim of this work. As we will soon see, this functional will be closely connected to the mechanism of emergent space in the AdS/CFT.

\subsection{Optimization of Vacuum States in 2D CFTs} \label{2dcftopt}
Let us first see how the optimization procedure works for vacuum states in 2D CFTs. We will study more general states later in later sections.

In 2D CFTs, we can always make the general metric into the diagonal form via a coordinate transformation. Thus the optimization is performed in the following ansatz:
\be
\begin{split}
& ds^2=e^{2\phi(z,x)}(dz^2+dx^2),  \\
& e^{2\phi(z=\ep,x)}=1/\ep^2,
\end{split}\label{met}
\ee
where the second condition specifies the boundary condition so that the discretization is fine-grained when we read off the wave function after the full path-integration. Obviously this is a special example of the ansatz (\ref{metand}). Thus the metric is characterized by the Weyl scaling function $\phi(z,x)$.

Remarkably, in 2D CFTs, we know how the wave function changes under such a local Weyl transformation.
Keeping the universal UV cut off $\ep$, the measure of the path-integrations of quantum fields in the CFT changes under the Weyl rescaling \cite{GM}:
\ba
[D\vp]_{g_{ab}=e^{2\phi}\delta_{ab}}=e^{S_L[\phi]-S_L[0]}\cdot [D\vp]_{g_{ab=\delta_{ab}}},
\ea
where $S_L[\phi]$ is the Liouville action\footnote{Here we take the reference metric is flat
 $ds^2=dz^2+dx^2$. Later in section (\ref{actionlii}), we will present the Liouville action
 for a more general reference metric.} \cite{Po} (see also \cite{GM,Wadia})
\be
S_L[\phi]=\frac{c}{24\pi}\int^\infty_{-\infty} dx \int^\infty_{\ep} dz \left[ (\de_x \phi)^2+(\de_z \phi)^2+\mu e^{2\phi}\right].
\ee
The constant $c$ is the central charge of the 2D CFT we consider. The kinetic term in $S_L$ represents the conformal anomaly and the potential term arises the UV regularization which manifestly breaks the Weyl invariance. In our treatment, we simply set $\mu=1$ below by suitable shift of $\phi$.

Therefore, the wave functional $\Psi_{g_{ab}=e^{2\phi}\delta_{ab}}(\ti{\vp}(x))$ obtained from the Euclidean path-integral for the metric (\ref{met}) is proportional to the one $\Psi_{g_{ab}=\delta_{ab}}(\ti{\vp}(x))$ for the flat metric (\ref{flat}) thanks to the conformal invariance. The proportionality coefficient is given by the Liouville action as follows\footnote{
Here we compare the optimized metric $g_{ab}=e^{2\phi}\delta_{ab}$ with $g_{ab}=\delta_{ab}$.
To be exact we need to take the latter to be the original one (\ref{flat}) i.e. $g_{ab}=\ep^{-2}\delta_{ab}$. However the different is just a constant factor multiplication and does not affect our arguments. So we simply ignore this.}
\ba
\Psi_{g_{ab}=e^{2\phi}\delta_{ab}}(\ti{\vp}(x))=e^{S_L[\phi]-S_L[0]}\cdot \Psi_{g_{ab}=\delta_{ab}}(\ti{\vp}(x)).
\ea

Let us turn to the optimization procedure. As proposed in \cite{CKMTW}, we argue that the optimization is equivalent to minimizing the normalization factor $e^{S_L[\phi]}$ of the wave functional, or equally the complexity functional $I_{\Psi_0}$ for the vacuum state $|\Psi_0\lb$ in 2D CFTs, can be identified as follows\footnote{In two dimensional CFTs, as we will explain in section \ref{actionlii}, due to the conformal anomaly we actually need to define a relative complexity by the difference of the Liouville action between two different metrics. However this does not change out argument in this section.}
\ba
I_{\Psi_0}[\phi(z,x)]=S_L[\phi(z,x)]. \label{compi}
\ea
The intuitive reason is that this factor is expected to be proportional to the number of repetition of the same operation (i.e. the path-integral in one site). In 2D CFTs, we believe this is only one quantity which we can come up with to measure the size of path-integration. Indeed it is proportional to the central charge, which characterizes the degrees of freedom.

Thus the optimization can be completed by requiring the equation of motion of Liouville action $S_L$ and this reads
\be
4\de_w \de_{\bar{w}}\phi=e^{2\phi},   \label{leom}
\ee
where we introduced $w=z+ix$ and $\bar{w}=z-ix$.

With the boundary condition $e^{2\phi(z=\ep,x)}=\ep^{-2}$, we can easily find the suitable solution to (\ref{leom}):
\be
e^{2\phi}=\frac{4}{(w+\bar{w})^2}=z^{-2},  \label{hyp}
\ee
which leads to the hyperbolic plane metric
\be
ds^2=\frac{dz^2+dx^2}{z^2}.
\ee
This justifies the heuristic argument to derive a hyperbolic plane H$_2$ in Fig.\ref{pathfig}.

Indeed, this hyperbolic metric is the minimum of $S_L$ with the boundary condition.  To see this, we rewrite
\be \label{newe}
\begin{split}
S_L=& \frac{c}{24\pi}\int dxdz \left[(\de_x\phi)^2+(\de_z\phi+e^{\phi})^2\right]
 -\frac{c}{12\pi}\int dx [e^{\phi}]^{z=\infty}_{z=\ep} ~\geq \frac{cL}{12\pi\ep},
\end{split}
\ee
where $L\equiv \int dx$ is the length of space direction and we assume the IR behavior
$e^{2\phi(z=\infty,x)}=0$. The final inequality in (\ref{newe}) is saturated if and only if
\be
\de_x\phi=\de_z\phi+e^{\phi}=0,
\ee
and this leads to the solution (\ref{hyp}).

In this way, we observe that the time slice of AdS$_3$ dual to the 2D CFT vacuum emerges after the optimization. We will see more evidences throughout this paper that geometries obtained from our optimization coincides with the time slice in AdS/CFT. This is consistent with the idea of tensor network description of AdS/CFT and can be regarded as its continuous version. We would like to emphasize that the above argument only depends on the central charge $c$ of the 2D CFT we consider.
Therefore this should be applied to both free and interacting CFTs including holographic ones.

It is also interesting to note that the optimized value of $S_L$, i.e. our complexity $C_{\Psi_0}$, scales linearly with respect to the momentum cut off $\ep^{-1}$ and the central charge $c$ as
\be
C_{\Psi_0}=\mbox{Min}_{\phi}[S_L[\phi]]= \frac{cL}{12\pi\ep},
\ee
and this qualitatively agrees with the behavior of the computational complexity \cite{SUR,Susskind} of a CFT ground state and the quantum information metric \cite{InfoM} for the same state,
both of which are given by the volume of time slice of AdS. In this relation, our minimization of $S_L$ nicely corresponds to the optimization of the quantum circuits which is
needed to define the complexity.

\subsection{Tensor Network Renormalization and Optimization}\label{tnre}

As argued in our preceding letter \cite{CKMTW} (see also \cite{MTW}), our identification of the Liouville action with a complexity i.e.(\ref{compi}) is partly motivated by an interesting connection between the tensor network renormalization (TNR) \cite{TNR} and our optimization procedure of Euclidean path-integral. This is because the number of tensors in TNR is an estimation of complexity and the Liouville action has a desired property in this sense, e.g. it is obvious that the Liouville potential term $\int e^{2\phi}$ (i.e. the volume) measures the number of unitary tensors in TNR. Soon later this argument was sharpened in the quite recent paper \cite{CzechC} where the number of isometries is argued to explain the kinetic term $\int (\de\phi)^2$ in Liouville theory.

\begin{figure}[t!]
  \centering
  \includegraphics[width=12cm]{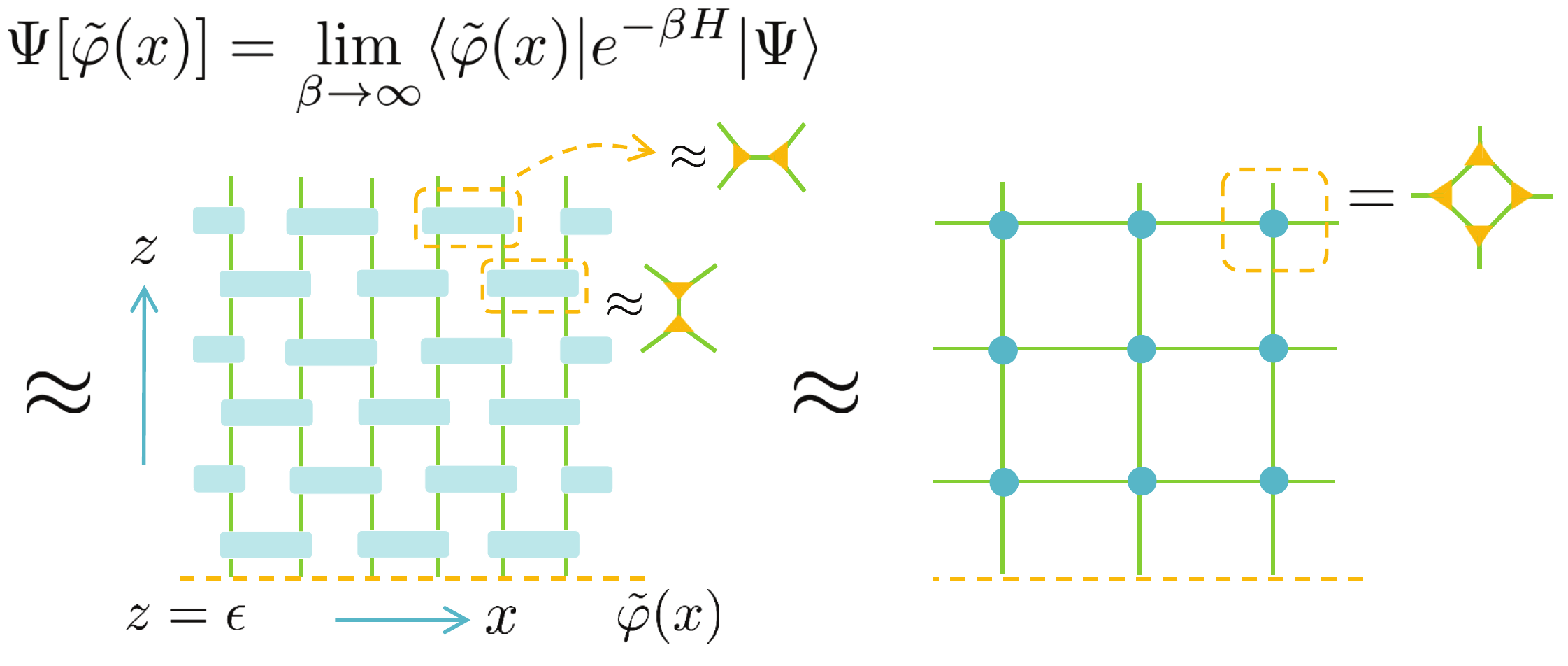}
\caption{The Euclidean path integral for the ground state wave functional $\Psi[\tilde{\varphi}(x)]$ can be approximately described by a tensor network on a square lattice.}
\label{fig:TNR1}
  \end{figure}

\begin{figure}[t!]
  \centering
  \includegraphics[width=12.5cm]{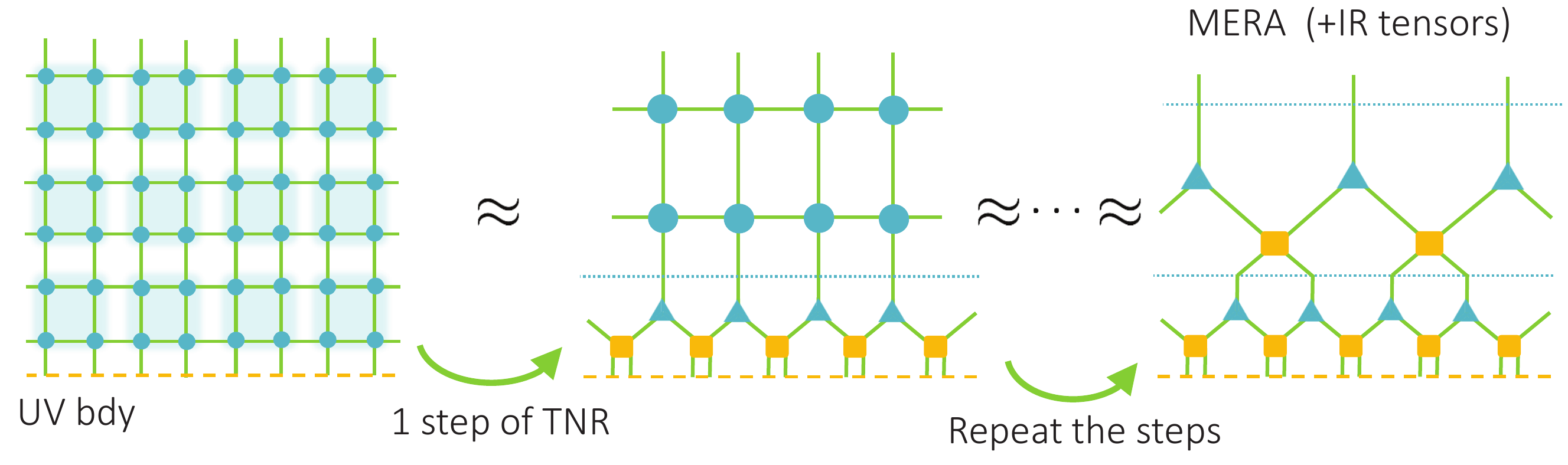}
  \caption{The tensor network renormalization (TNR) gradually makes the coarse-grained tensor network with removing short-range entanglement.
From the UV boundary, isometries (coarse-graining) and unitaries (disentanglers) accumulate and the MERA network grows with the TNR steps.}
\label{fig:TNR2}
  \end{figure}

An Euclidean path-integral on a semi-infinite plane (or cylinder) with a boundary condition on the edge gives us a ground state wave functional
in a quantum system. The path-integral can be approximately described by a tensor network which is a collection of tensors contracted with each other. Using the Suzuki-Torotter decomposition \cite{STdecomp} and the singular value decomposition of the tensors, we can rewrite the Euclidean path-integral into a tensor network on a square lattice (Fig.\ref{fig:TNR1}).
Tensor network renormalization (TNR) is a procedure to reorganize the tensors to ones on a coarser lattice by inserting projectors (isometries) and unitaries (disentanglers) with removing short-range entanglement.\footnote{Note that by adding a dummy or ancilla state $|0\lb$ we can equivalently regard an isometry as a unitary.} This is a step of TNR (Fig.\ref{fig:TNR2}).
Repeating this procedure, we can generate a RG flow properly and end up with a tensor network at the IR fixed point.
For the ground state wave functional in a CFT, it ends up with a MERA (Multi-scale Entanglement Renormalization Ansatz) network made of isometries and disentanglers.
The MERA network clearly contains smaller numbers of the tensors than ones in the tensor network on the original square lattice before the coarse-graing.
In this sense, this MERA network is an optimal tensor network to approximately describe the Euclidean path-integral.

Our optimization procedure is motivated by TNR.
In our procedure (Fig.\ref{pathfig}), the tensor network on the square lattice corresponds to the Euclidean path-integral on flat space with a UV cutoff $\epsilon$.
Changing the tensor network with inserting isometries and entanglers corresponds to deforming the back-ground metric for the path-integral.
And the MERA network, which is the tensor network after the TNR procedure, approximately corresponds to the optimized path-integral.
  
Actually, it is not difficult to estimate the amount of complexity for each tensor network during the TNR optimization procedure, by identifying the complexity with the number of tensors, both isometries (coarse-graining) and unitaries (disentanglers).
For simplicity, consider an Euclidean path-integral for the ground state wave function in a 2d CFT, which is performed on the upper half plane
($\epsilon < z < \infty, -\infty < x <+\infty$).
First we consider the original square lattice. Since we suppose that each tensor have unit area,
the uniform metric is given by $e^{2\phi(z)}=\ep^{-2}$ as in (\ref{flat}). Therefore, the total number of tensors, which are only unitaries, is estimated from the total volume:
\be
\int_{-\infty}^{\infty} dx \int_{\epsilon}^{\infty} dz \frac{1}{\epsilon^2}
= \int_{-\infty}^{\infty} dx \int_{\epsilon}^{\infty} dz e^{2\phi}.
\ee

Then, performing the TNR procedure, the number of the tensors or the square lattice sites is reduced by the factor $(1/2)^2$ per step.
On the other hand, the isometries and disentanglers accumulate from the UV boundary \cite{TNR}.
Refer to Fig.\ref{fig:TNR2}.

\begin{figure}[b!]
  \centering
  \includegraphics[width=8.5cm]{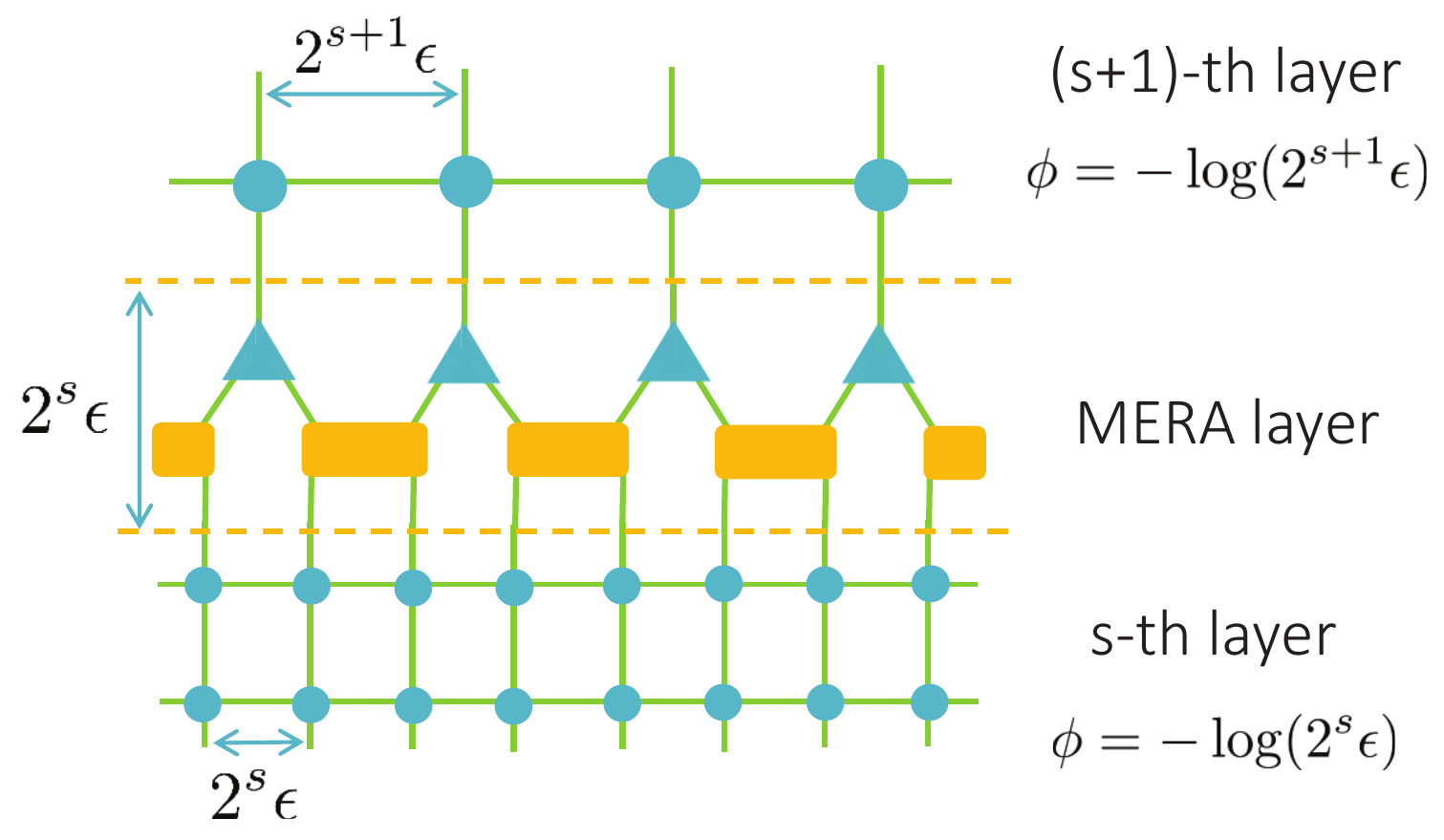}
  \caption{The tensor network produced when we have a shift of $\phi$ at a specific layer.
This also represents the one step ($s$-th) contribution in the process of tensor network renormalization, which finally reaches the MERA network. This corresponds to $s-$th terms
$\int^{2^s \epsilon}_{2^{s-1} \epsilon} dz(\ddd)$ in (\ref{tnrcomp}).}
\label{fig:TNR3}
  \end{figure}

At the $k$-th step of TNR, the total area changes into
\be
\int_{-\infty}^{\infty} dx \int^{\infty}_{2^{k} \epsilon} dz  \frac{1}{(2^k \epsilon)^2} + \sum_{s = 1}^{k} \int_{-\infty}^{\infty} dx \int^{2^s \epsilon}_{2^{s-1} \epsilon} dz \left( \frac{1}{(2^{s-1}\epsilon) \cdot (2^{s}\epsilon)} + \frac{1}{(2^{s}\epsilon)^2} \right).
\label{tnrcomp}
\ee
The first term is the contribution from the tensors on the coarser lattice.
The second term is the contribution from the MERA network.
For the $s$-th layer of the MERA network, we have $dx dz/( (2^{s-1}\epsilon) \cdot (2^{s}\epsilon) )$ isometries and $dx dz/(2^{s}\epsilon)^2$ per unit cell. This contribution is depicted in
Fig.\ref{fig:TNR3}.
\\
This network corresponds to the metric
\be
e^{2\phi}=
\Bigl\{
\begin{array}{c}
  (2^{k}\ep)^{-2}\ \ \ (z\geq 2^{k}\ep). \\
  z^{-2} \ \ \ (z<2^{k}\ep).
\end{array}
\ee
Obviously, the first and third term in (\ref{tnrcomp}) are approximated by the Liouville potential integral $\int e^{2\phi}$ \cite{CKMTW}. The second term arises because of the non-zero gradient
 of $\phi$ and is estimated by the kinetic term $\int (\de\phi)^2$ \cite{CzechC}.

\section{Optimizing Various States in 2D CFTs}\label{Optimization2D}
Here we would like to explore optimizations in 2D CFTs for more general quantum states. First it is useful to remember that the general solutions to the Liouville equation (\ref{leom}) is well-known (see e.g.\cite{Se,GM}):
\be
e^{2\phi}=\frac{4A'(w)B'(\bar{w})}{(1-A(w)B(\bar{w}))^2}.
\ee
Note that functions $A(w)$ and $B(\bar{w})$ describe the degrees of freedom of conformal mappings.
For example, if we choose
\be
A(w)=w,\ \  B(\bar{w})=-1/\bar{w},  \label{abf}
\ee
then we reproduce the solution for vacuums states (\ref{hyp}).

\subsection{Finite Temperature States}\label{opt:ft}
Consider a 2D CFT state at a finite temperature $T=1/\beta$. In the thermofield double description \cite{MaE}, the wave functional is expressed by an Euclidean path-integral on a strip defined by $-{\beta \over 4}(\equiv z_{1})< z< {\beta \over 4}(\equiv z_2)$ in the Euclidean time direction, more explicitly
\be\label{tfdwfr}
\begin{split}
 \Psi[\ti{\vp}_1(x),\ti{\vp}_2(x)]&=\int \left(\prod_{x}\prod_{-{\beta \over 4}<z<{\beta \over 4}}D\vp(z,x)\right)e^{-S_{CFT}(\vp)} \\
& \times\!\!\prod_{-\infty<x<\infty}\!\!\!\delta\big(\vp\left(z_1,x\right)\!-\!\ti{\vp}_1(x)\big)
~\delta\big(\vp\left(z_2,x\right)\!-\!\ti{\vp}_2(x)\big).
\end{split}
\ee
where $\ti{\vp}_1(x)$ and $\ti{\vp}_2(x)$ are the boundary values for the fields of the CFT (i.e. $\ti{\vp}(x)$) at $z= \mp {\beta \over 4}$ respectively.\\
Minimizing the Liouville action $S_L$ leads to the solution in (\ref{abf}) given by:
\be
A(w)=e^{{2 \pi i w\over \beta}},\ \  B(\bar{w})=-e^{{2 \pi i \bar{w} \over \beta}}.
\ee
This leads to
\be \label{phforbtz}
e^{2\phi}=\frac{16 \pi^2}{\beta^2}\frac{e^{{2 \pi i \over \beta}(w+\bar{w})}}{\left(1+ e^{{2 \pi i \over \beta}(w+\bar{w})}\right)^2}=\frac{4 \pi^2}{\beta^2} \sec^2 \left({2 \pi z\over \beta}\right).
\ee
If we perform the following coordinate transformation
\be
\tan \left({ \pi z\over \beta}\right) = \tanh \left({ \rho \over 2}\right),
\ee
then we obtain the metric
\be \label{BTZmet2}
ds^2 = d\rho^2 + \frac{4 \pi^2}{\beta^2}~ \cosh^2\rho~ dx^2,
\ee
which coincides with the time slice of eternal BTZ black hole  (i.e. the Einstein-Rosen bridge) \cite{MaE}. 

\subsection{CFT on a Cylinder and Primary States}\label{opt:pr}

 Now consider 2D CFTs on a cylinder (with the circumference $2\pi$), where the wave functional is defined on a circle $|w|=1$ at a fixed Euclidean time. After the optimization procedure, we obtain the geometry $A(w)=w$ and $B(\bar{w})=\bar{w}$ given by
 \be
 e^{2\phi(w,\bar{w})}=\frac{4}{(1-|w|^{2})^2}, \label{pdiskm}
 \ee
 which is precisely the Poincare disk and is the solution to (\ref{leom}).

Then we consider an excited state given by a primary state $|\ap\lb$. This is created by acting a primary operator $O_\ap(w,\bar{w})$ with the conformal dimension $h_\ap=\bar{h}_\ap$ at the center $w=\bar{w}=0$. Its behavior under the Weyl re-scaling is expressed as
\be
O(w,\bar{w})\propto e^{-2h_\ap \phi}.
\ee
 Thus the dependence of the wave function on $\phi$ looks like
\ba
\Psi_{g_{ab}=e^{2\phi}\delta_{ab}}(\ti{\vp})\simeq e^{S_L}\cdot e^{-2h_\ap\phi(0)}\cdot \Psi_{g_{ab}=\delta_{ab}}(\ti{\vp}).
\ea
This shows that the complexity function should be taken to be
\be
I_{\ap}[\phi(w,\bar{w})]=S_L[\phi(w,\bar{w})]-2h_\ap\phi(0).
\ee
The equation of motion of $I_\ap$ leads to
\be
4\de_w \de_{\bar{w}}\phi-e^{2\phi}+2\pi(1-a)\delta^2(w)=0
\label{eomf},
\ee
where we set
\be
a=1-\frac{12h_\ap}{c}.  \label{dila}
\ee
The solution can be found as
\be
A(w)=w^a,\ \ \ B(\bar{w})=\bar{w}^a,
\ee
which leads to the expression:
\be
e^{2\phi}=\frac{4a^2}{|w|^{2(1-a)}(1-|w|^{2a})^2}. \label{defcit}
\ee
Since the angle of $w$ coordinate is $2\pi$ periodic, this geometry has the deficit angle $2\pi(1-a)$.

 Now we compare this geometry with the time slice of the gravity dual predicted from AdS$_3/$CFT$_2$. It is given by the conical deficit angle geometry (\ref{defcit}) with the identification
\be
a=\s{1-\frac{24h_\ap}{c}}.  \label{three}
\ee
Thus, the geometry from our optimization (\ref{dila}) agrees with the gravity dual (\ref{three}) up to the first order correction when $h_{\alpha}\ll c$, i.e. the case where the back-reaction due to the point particle is very small.

It is intriguing to note that if we consider the quantum Liouville theory rather than the classical one, we find the perfect matching. In the quantum Liouville theory, we introduce a parameter $\gamma$ such that $c=1+3Q^2$ and $Q\equiv\frac{2}{\gamma}+\gamma$. The chiral conformal dimension of the primary operator $e^{\frac{2\beta}{\gamma}\phi}$ is given by $\frac{\beta(Q-\beta)}{2}$. If the central charge is very large so that the 2D CFT has a classical gravity dual, we find
\be
a\simeq 1-\beta\gamma\simeq \s{1-\frac{24h_\ap}{c}},
\ee
 which indeed agrees with the gravity dual (\ref{three}) even when $h_\ap/c$ is finite.

This agreement may suggest that the actual optimized wave functional is given by a `quantum' optimization defined as follows:
\ba
\Psi_{\mbox{opt}}[\ti{\vp}]=\left[\int D\phi(x,z)e^{-S_L[\phi]}\left(\Psi_{g_{ab}=\delta_{ab}}[\ti{\vp}]\right)^{-1}\right]^{-1}.
\label{qwe}
\ea
If we take the semi-classical approximation when $c$ is large, we reproduce our classical optimization. It is an important future problem to understand the exact for of the proposal at the quantum level.

\subsection{Liouville Equation and 3D AdS Gravity}

In the above we have seen that the minimizations of Liouville action, which corresponds to the
optimization of Euclidean path-integrals in CFTs, lead to hyperbolic metrics which fit nicely with canonical time slices of bulk AdS in various setups of AdS$_3/$CFT$_2$. If this derivation of time slice metric in AdS$_3$ really explains the mechanism of emergence of AdS in AdS/CFT, it should
fit nicely with the dynamics of AdS gravity for the whole 3D space-time.
One natural coordinate system in 3D gravity for our argument is as follows
\be
ds^2=R_{AdS}^2\left(d\rho^2+\cosh^2\rho\cdot e^{2\phi}dyd\bar{y}\right). \label{ourm}
\ee
Indeed the Einstein equation $R_{\mu\nu}+\frac{2}{R_{AdS}^2}g_{\mu\nu}=0$ is equivalent to the equation of motion in the Liouville theory: $4\de_y\de_{\bar{y}}\phi=e^{2\phi}$.

It is also useful to remember that connections between Liouville theory and 3D AdS gravity were discussed in earlier papers \cite{TV,CaP,CHD,CS1,CS2,EH1,EH2,EH3} (refer to \cite{CaPreview} for a review). Especially the direct connection between the equation of motion in the SL$(2,R)$ Chern-Simons gauge theory description of AdS gravity \cite{WittenCS} and that of Liouville theory was found in \cite{CHD} (see also closely related arguments \cite{CS1,CS2,EH1,EH2}).

Indeed, we can find a coordinate transformation which maps the metric (\ref{ourm}) into the one from \cite{CHD}, where the map gets trivial only in the near boundary limit $\rho\to\infty$. This shows that we can identify these two appearances of Liouville theory from 3D AdS gravity by a non-trivial bulk coordinate transformation. We presented the details of this transformation in the appendix \ref{ap:3dgravity}.

Notice also that we did not fix the overall normalization of the optimized metric or equally the AdS radius $R_{AdS}$ because in our formulation it depends on the precise definition of UV cut off. However, we can apply the argument of \cite{MTW} for the symmetric orbifold CFTs and can heuristically argue that $R_{AdS}$ is proportional to the central charge $c$. This is deeply connected to the fact that we find the sub AdS scale locality in gravity duals of holographic CFTs.

\section{Reduced Density Matrices and EE}\label{HEE2DCFT}

\begin{figure}[t!]
  \centering
  \includegraphics[width=7cm]{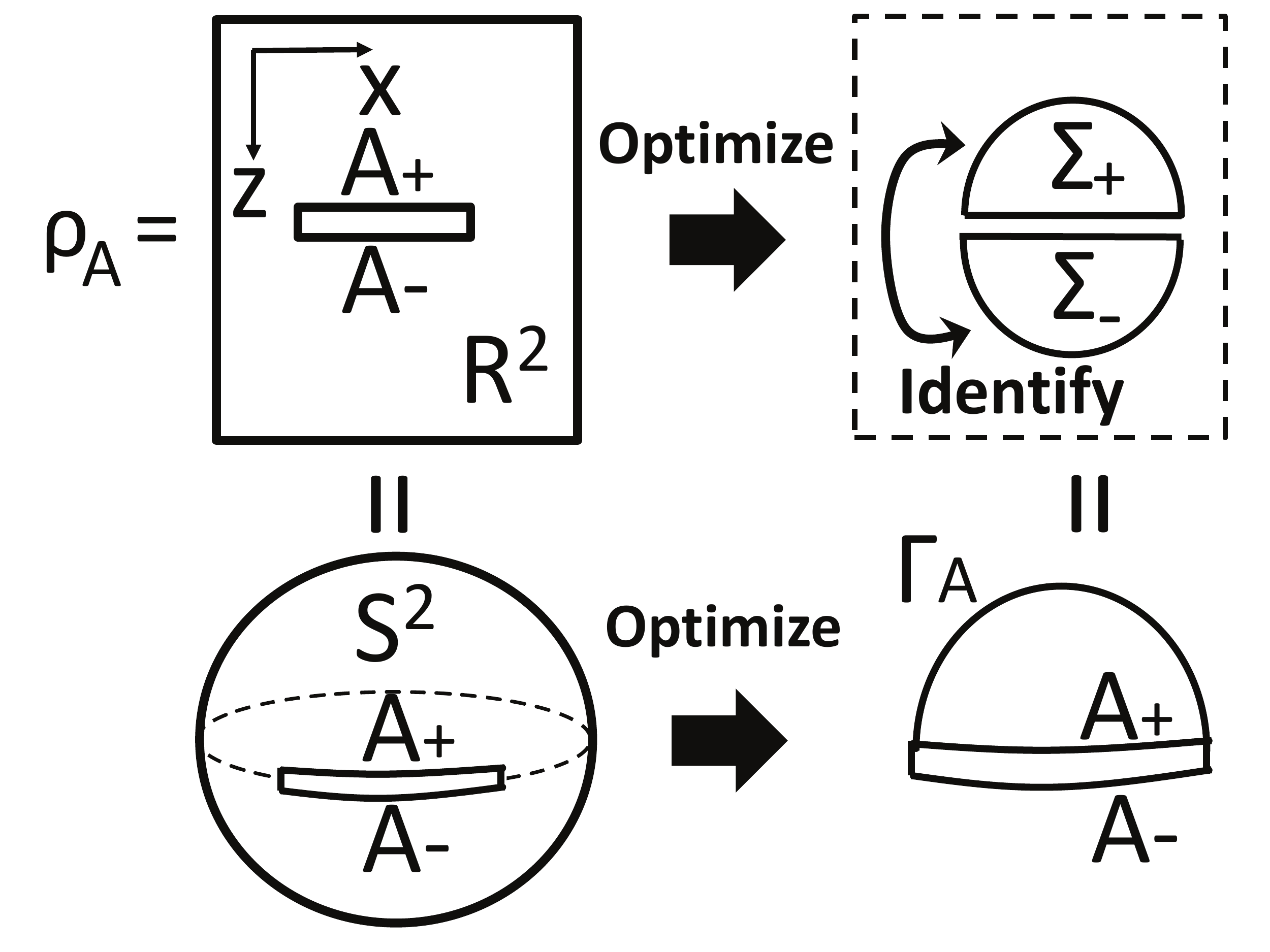}
  \caption{The optimization of path-integral for a reduced density matrix. The upper left picture is the definition of $\rho_A$ in terms of the path-integral in flat space. This is conformally mapped into a sphere with a open cut depicted in the lower left picture. The upper right one is the one after the optimization and is equivalent to a geometry which is obtained by pasting two identical entanglement wedges along the geodesic (=the half circle) as shown in the lower right picture.}
\label{redfig}
  \end{figure}

Consider an optimization of path-integral representation of reduced density matrix $\rho_A$ in a two dimensional CFT defined on a plane R$^2$. We simply choose the subsystem $A$ to be an interval $-l\leq x\leq  l$ at $z(=-\tau)=\ep$. $\rho_A$ is defined from the CFT vacuum by tracing out the complement of $A$ (the upper left picture in Fig.\ref{redfig}).
  
\subsection{Optimizing Reduced Density Matrices}

The optimization procedure is performed by changing the background metric as in (\ref{met}), where the boundary condition of $\phi$ is imposed around the upper and lower edge of the slit $A$. Refer to Fig.\ref{redfig} for a sketch of this procedure. The plane R$^2$ is conformally mapped into a sphere S$^2$. Therefore
the optimization is done by shrinking the sphere with an open cut down to a much smaller one so that the Liouville action is minimized.

To make the analysis clear, let us divide the final manifold into two halves by cutting along the horizontal line $z=0$, denoted by $\Sigma_{+}$ and $\Sigma_{-}$. The boundary of $\Sigma_{\pm}$ consist of two parts:
\be
\de \Sigma_{\pm}=A_{\pm}\cup \Gamma_{A},
\ee
where $\Gamma_A$ in both $\de \Sigma_{+}$ and $\de \Sigma_{-}$ are identified so that the topology of the final optimized manifold $\Sigma_+\cup \Sigma_{-}$ is a disk with the boundary
$A_{+}\cup A_{-}$. On the boundary $A_{+}\cup A_{-}$ we have $e^{2\phi}=1/\ep^2$.

The optimization of each of $\Sigma_{\pm}$ is done by minimizing the Liouville action with boundary contributions. The boundary action in the Liouville theory \cite{FZZ} reads
\be
S_{Lb}=\frac{c}{12\pi}\int_{\de\Sigma_{\pm}} ds [K_0\phi+\mu_B e^{\phi}],  \label{bLv}
\ee
where $K_0$ is the (trace of) extrinsic curvature of the boundary $\de\Sigma_{\pm}$ in the flat space. If we describe the boundary by $x=f(z)$, then the extrinsic curvature in the flat
metric $ds^2=dz^2+dx^2$, is given by $K_0=-\frac{f''}{(1+(f')^2)^{3/2}}$. On the other hand, the final term is the boundary Liouville potential. Since $\Sigma_+$ and $\Sigma_{-}$ are pasted along the boundary smoothly, we set $\mu_B=0$ for our $\rho_A$ optimization.\footnote{Non-zero $\mu_B$ leads to a jump of the extrinsic curvature which will be used later.}

Now, to satisfy the equation of motion at the boundary $\Gamma_A$, we impose the Neumann boundary condition\footnote{On the cuts $A_{\pm}$ we imposed the Dirichlet boundary condition. The reason
why we imposed the Neumann one on $\Gamma_A$ is simply because the manifold is smoothly connected to the other side on $\Gamma_A$.} of $\phi$. This condition (when $\mu_B=0$) is explicitly written as
\be
(n^x\de_x+n^z\de_z)\phi+K_0=0. \label{libc}
\ee
where $n^{x,z}$ is the unit vector normal to the boundary in the flat space. Actually this is simply expressed as $K=0$, where $K$ is the extrinsic curvature in the curved metric (\ref{met}).
This fact can be shown as follows. Consider a boundary $x=f(z)$ in the two dimensional space defined by the metric $ds^2=e^{2\phi(z,x)}(dz^2+dx^2)$. The out-going normal unit vector $N^a$ is given by
\be
 N^z=e^{-\phi(z,x)}n^z=\frac{-f'(z)e^{-\phi(z,x)}}{\s{1+f'(z)^2}}, ~~N^x=e^{-\phi(z,x)}n^x=\frac{e^{-\phi(z,x)}}{\s{1+f'(z)^2}},
\ee
where $n^a$ is the normal unit vector in the flat space $ds^2=dz^2+dx^2$.
The extrinsic curvature (=its trace part) at the boundary is defined by $K=h^{ab}\nabla_a N_b$, where all components are projected to the boundary whose induced metric is written as $h_{ab}$.
Explicitly we can calculate $K$ as follows:
\ba
K=\frac{e^{-\phi(z,x)}}{\s{1+f'(z)^2}}\left[\de_x\phi-f'\de_z\phi-\f{f''}{1+(f')^2}\right]
=e^{-\phi(z,x)}\left[n^a\de_a\phi+K_0\right].
\ea

In this way, the Neumann boundary condition requires that the curve $\Gamma_A$ is geodesic. By taking the bulk solution given by the hyperbolic space $\phi=-\log z+$const., the geodesic $\Gamma_A$ is given by the half circle $z^2+x^2=l^2$. Thus, this geometry obtained from the optimization of $\rho_A$, coincides with (two copies of) the entanglement wedge \cite{RT,HRT,EW1,EW2,EW3}.

Note that if we act a local operator inside the entanglement wedge in the original flat space, then this excitation survives after the optimization procedure. However, if we act the operator outside, then the excitation is washed out under the optimization procedure and does not reflect the reduced density matrix $\rho_A$ as long as we neglect its back-reaction.

\subsection{Entanglement Entropy}

\begin{figure}[h!]
  \centering
  \includegraphics[width=7cm]{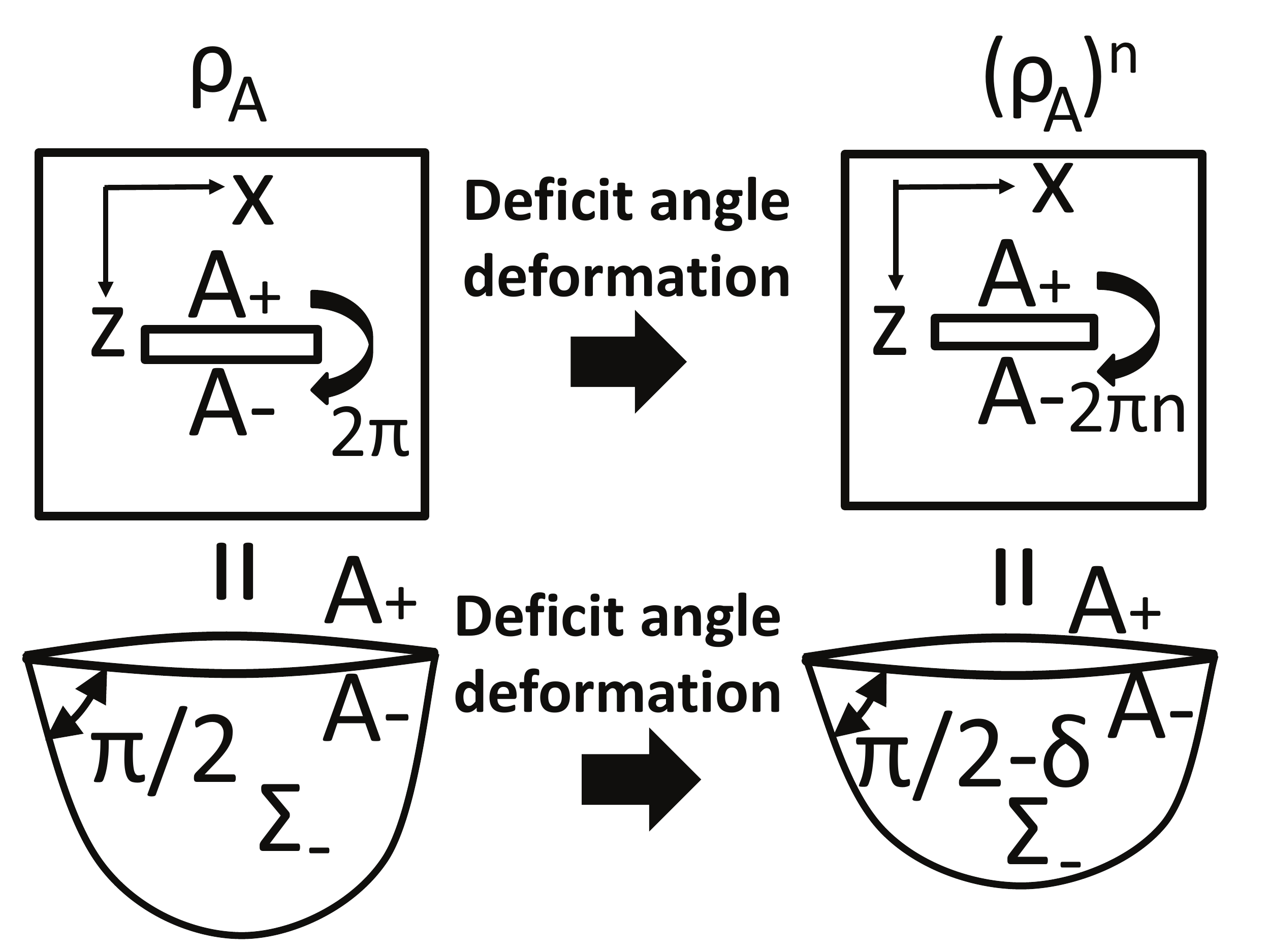}
  \caption{The computation of entanglement entropy using the optimized reduced density matrix. Following the replica method shown in the upper pictures, we consider the evaluation of $\rho_A^n$. We assume the analytical continuation such that $n$ is very close to $1$ such that
  $\delta\equiv \pi (1-n)\ll 1$. Thus this describes an infinitesimally small (negative) deficit angle deformation. After the optimization, we obtain the conical geometry in the lower right picture with $\delta=\pi (1-n)$.}
\label{figent}
  \end{figure}

Next we evaluate the entanglement entropy by the replica method. Consider an optimization of the matrix product $\rho_A^n$.
We assume an analytical continuation of $n$ with $|n-1|\ll 1$.
The standard replica method leads to a conical deficit angle $2\pi(1-n)\equiv 2\delta$
at the two end points of the interval $A$. Thus, after the optimization, we get a geometry with the corner angle $\pi/2+\pi(n-1)$ instead of $\pi/2$ (the lower right picture in Fig. \ref{figent}). This modification of the boundary $\Gamma_A$ is equivalent to shifting the extrinsic curvature from $K=0$ to $K=\pi(n-1)$. Indeed, if we consider the boundary given by $x^2+(z-z_0)^2=l^2$, we get $K=z_0/l$. When $z_0$ is infinitesimally small, we get $x\simeq l+(z_0/l)\cdot z+O(z^2)$ near the boundary point $(z,x)=(0,l)$. Therefore the corner angle is shifted to be $\pi/2-\delta$ with $\delta\simeq -z_0/l$ (for the definition of $\delta$, refer also to lower pictures in Fig.\ref{figent}). Therefore we find the relation $K\simeq -\delta$.
If we set the boundary Liouville term in (\ref{bLv}) non-zero $\mu_B\neq 0$,
the boundary condition
is modified from (\ref{libc}) i.e. $K=0$ into $K+\mu_B=0$. Thus the desired angle shift (or negative deficit angle) is realized by setting $\mu_B=\pi(1-n)$. In the presence of infinitesimally small $\mu_B$ we can evaluate the Liouville action by a probe approximation neglecting all back reactions. By taking a derivative with respect to $n$, we obtain the entanglement entropy\footnote{The abuse of notation for the entanglement entropy and the Liouville aciton should be clear form the context.} $S_A$:
\be
S_A=-\de_{n}\left[2\times\frac{c\mu_B}{12\pi}\int_{\Gamma_A}ds\ e^{\phi}\right]_{n=1}= \frac{c}{6}\int_{\de\Sigma_{+}}ds\ e^{\phi}=\frac{c}{3}\log\frac{l}{\ep},\label{entd}
\ee
reproducing the well-known result \cite{HLW}. The lower left expression (\ref{entd}) $\frac{c}{6}\int_{\de\Sigma_{+}} e^\phi$
precisely agrees with the holographic entanglement entropy formula \cite{RT,HRT} as $\Gamma_A$ has to be the geodesic due to the boundary condition.

\subsection{Subregion Complexity}
Finally we would like to evaluate the value of Liouville action $S_L[\phi]$ in the reduced sub-region. It is natural to argue that this provides a definition of complexity for the reduced density matrix $\rho_A$. For various earlier proposals for holographic subregion complexity refer to \cite{Alis,CMR}.

As in the previous section we take $A$ to be the interval $-l\leq x\leq l$. By computing the action for two copies of the half disk $x^2+z^2\leq l^2$ with the solution $\phi=-\log z$, we find
\be
\begin{split}
S_L=&\frac{c}{12\pi}\int dxdz \left[(\de_z\phi)^2+e^{2\phi}\right]+\frac{c}{6\pi}\int ds K_0\phi \\ =& \frac{c}{6\pi}
\int^l_{\ep}dz\frac{2\s{l^2-z^2}}{z^2}+\frac{c}{6\pi}\int^{\pi/2}_{-\pi/2}[-\log(l\cos\theta)] \\
=& \frac{c}{6\pi}\left[ \frac{2l}{\ep}-\pi-\pi\log\left(\frac{l}{2}\right)\right].
\end{split}
\ee
It will be interesting to compute and explore it further for more general states and we leave it as an open future problem.

\section{Energy Momentum Tensor in 2D CFTs}\label{EMTensor}
One of the most fundamental objects in two dimensional CFTs is the energy momentum tensor and in this section we show how to extract it from our optimization. Since we already know how to compute entanglement entropy, our derivation will be based on the first-law of entanglement that relates changes in entanglement entropy of an interval to the energy momentum tensor. More precisely, as shown in \cite{Bhattacharya:2012mi}, under small perturbations of a quantum state, the change of entanglement entropy of a small interval $A=[-l/2,l/2]$ is proportional to $T_{tt}$
\be
\Delta S_{A}\simeq \frac{\pi l^2}{3}T_{tt}.
\ee
On the other hand, in our approach, the change in entanglement entropy under a small variation of a quantum state is captured by the variation in the Liouville field $\phi(z)\to \phi_0(z)+\delta \phi(z)$. Moreover, for small perturbations we can write
\be
\delta \phi(z)=\f{z^2}{2}\partial^2_z\delta\phi(z)+O(z^4),
\ee
such that the change in entanglement entropy in perturbed state becomes
\ba
\Delta S_A \simeq\frac{c}{6}\int  \delta\phi\cdot e^{\phi} ds \simeq \frac{c}{6}\de_z^2\delta\phi\int^{l/2}_0 \frac{z\,dz}{\s{1-4z^2/l^2}}=\frac{cl^2}{24}\de_z^2\delta\phi.
\ea
Comparing with the first law, we can now match the energy momentum tensor
\be
T_{tt}=\frac{c}{8\pi}\de_z^2\delta\phi.
\ee
Let us now compare this with our explicit examples. The vacuum solution is given by $\phi_0(z)=-\log\left(z\right)$. Then, after a simple shift, the thermofield double solution \eqref{phforbtz} can be written as
\be
\phi(z)=-\log\left(\f{\beta}{2\pi}\sin\left(\frac{2\pi z}{\beta}\right)\right)\simeq\phi_0(z)+\f{2\pi^2}{3\beta^2}z^2+O(z^4)
\ee
and we obtain the well known result
\be
T_{tt}=\f{\pi c}{6\beta^2}.
\ee
Similarly, writing our conical singularity solution  \eqref{defcit} in coordinates $w=\exp(z+ix)$ and $\bar{w}=\exp(z-ix)$, we get
\be
\phi(z)=-\log\left(\f{1}{a}\sinh\left(a z\right)\right)\simeq \phi_0(z)-\f{a^2z^2}{6}+O(z^4),
\ee
and the known energy momentum tensor
\be
T_{tt}=-\f{a^2c}{24\pi},
\ee
that for $a=1$ reproduces the Casimir energy.

Let us also point the interesting consistency of the above result with the Liouville energy momentum tensor. Namely, it is well known that by varying the action with respect to the background "reference" metric one can derive the Liouville energy momentum tensor. The corresponding holomorphic and anti-holomorphic classical energy momentum tensors are
\bea
T(w)&=&\f{c}{12\pi}\left(\partial^2_w\phi-\left(\partial_w \phi\right)^2\right),\\
\bar{T}(\bar{w})&=&\f{c}{12\pi}\left(\partial^2_{\bar{w}}\phi-\left(\partial_{\bar{w}} \phi\right)^2\right).
\eea
One can check that, for our solutions, these energy momentum tensors match the ones computed form the first law. In general we can use the first law for entanglement entropy in states conformally mapped to the vacuum (see e.g. \cite{Sheikh-Jabbari:2016znt,BHHM}) and show that the increase in the entropy is proportional to the (constant) Liouville energy momentum tensor.

\section{Evaluation of $S_L$ in 2D CFTs}\label{actionlii}

Here we first explain the properties of Liouville action $S_L$ in general setups with boundaries.
We will find that it depends on the reference metric and it does not seem to be possible to
define its absolute value, which is due to the conformal anomaly in 2D CFTs. Rather we are lead to introduce an functional defined by a difference of Liouville action denoted by $I_L[g_2,g_1]$,
where $g_1$ is the reference metric and $g_2$ is the final metric after the optimization.
$I_L[g_2,g_1]$ is expected to measure of the complexity between the two path-integrals in $g_1$
and $g_2$. Having them in mind, we proceed to explicit evaluations of $I_L[g_1,g_2]$ in various cases.

\subsection{General properties of the Liouville action}\label{weyl2d}

We start with a two dimensional space $\mathcal{M}$ described by the metric $ds^2=g_{ab}dx^a dx^b$ ($a,b=1,2$), which is called the reference metric. We now perform the Weyl transformation and define
the rescaled metric:
\be \label{mettot}
ds^2 = e^{2\phi}g_{ab}dx^adx^b.
\ee
The Liouville action corresponding to this Weyl rescaling is given by
\be \label{defSL}
S_L[\phi,g_{ab}] = \frac{c}{24\pi}\int_\mathcal{M} d^2x \sqrt{g} \bigg[g^{ab} \partial_a \phi \partial_b \phi+e^{2\phi}+R_g \phi \bigg]+\frac{c}{12\pi}\int_{\partial \mathcal{M}}ds \sqrt{h} K_g \phi,
\ee
where
\be \nonumber
\begin{split}
 \mathcal{M} =~ &\text{The 2-dim manifold with co-ordinates}~ \{x,y\} \\
 \partial \mathcal{M}  =~ &\text{The boundary of} ~\mathcal{M}~\text{with the coordinate} ~s \\
f(x,y) = ~ &\text{The equation for the boundary}~\partial \mathcal{M} \\
R_g  =~ &\text{Ricci scalar for the metric}~g_{ab} \\
n_a = ~& \pm {\partial_a f \over \sqrt{g^{bc}\partial_b f \partial_c f} }= \text{The unit normal to the boundary }~\partial \mathcal{M} \\
h_{ab} =~& g_{ab} -n_a n_b =\text{Induced metric on} ~\partial \mathcal{M}, ~~\text{such that} ~h^{ab}n_b=0, \\
K_g = ~& g^{ab} \nabla_a n_b = \text{Trace of the extrinsic curvature of} ~\partial \mathcal{M}
\end{split}
\ee

Now let us consider the following transformation parameterized by the function $A(x,y)$
\be \label{trans}
\begin{split}
\phi(x,y) \rightarrow \tilde{\phi}(x,y) = \phi(x,y)-A(x,y) \\
g_{ab} (x,y) \rightarrow \tilde{g}_{ab}(x,y) = e^{2A(x,y)}g_{ab}(x,y)
\end{split}
\ee
such that the final metric in \eqref{mettot} is invariant.

Let us note the following relations\footnote{For deriving the last relation
in \eqref{relations} we used
\be
\begin{split}
\tilde{n}_a = ~& \pm {\partial_a f \over \sqrt{\tilde{g}^{bc}\partial_b f \partial_c f} } = \pm e^{-A} {\partial_a f \over \sqrt{g^{bc}\partial_b f \partial_c f} } = e^{-A} n_a, \end{split}
\ee
and also
\be
\begin{split}
\tilde{K}_{\tilde{g}} = \tilde{g}^{ab} \tilde{\nabla}_a \tilde{n}_b = {1 \over \sqrt{\tilde{g}}} \partial_a (\sqrt{\tilde{g}}~ \tilde{n}^a) = {e^{2A} \over \sqrt{g}} \partial_a (e^A \sqrt{g}~ n^a) = e^{-A} [ K_g +n^a \partial_a A]
\end{split}
\ee
}
\be \label{relations}
\begin{split}
\sqrt{g} =&e^{2A}\sqrt{\tilde{g}},~~ \tilde{n}_a = e^A n_a,~~\sqrt{h} = e^{A}\sqrt{\tilde{h}}, \\
\tilde{R}_{\tilde{g}} =& e^{-2A} [R_g - 2 \nabla^2 A], ~~
\tilde{K}_{\tilde{g}} =e^{-A} [ K_g +n^a \partial_a A]
\end{split}
\ee

Therefore the transformed Liouville action becomes
\be
\begin{split}
S_L[\tilde{\phi},\tilde{g}_{ab}] =&\frac{c}{24\pi} \int_\mathcal{M} d^2x \sqrt{\tilde{g}} \bigg[ \tilde{g}^{ab}~ \partial_a \tilde{\phi}~ \partial_b \tilde{\phi}+ e^{2\tilde{\phi}}+
\tilde{R}_{\tilde{g}}~ \tilde{\phi} \bigg] \\
&+\frac{c}{12\pi} \int_{\partial \mathcal{M}}ds \sqrt{\tilde{h}}~ \tilde{K}_{\tilde{g}}~  \tilde{\phi}
\end{split}
\ee
Using the relations in \eqref{relations} it can be checked that
\be \label{term1}
\begin{split}
& \int_\mathcal{M} d^2x \sqrt{\tilde{g}} \bigg[ \tilde{g}^{ab}~ \partial_a \tilde{\phi}~ \partial_b \tilde{\phi}+ e^{2\tilde{\phi}}+
\tilde{R}_{\tilde{g}}~ \tilde{\phi} \bigg] \\ & ~~ = \int_\mathcal{M} d^2x \sqrt{g} \bigg[g^{ab}~ \partial_a \phi~ \partial_b \phi + e^{2 \phi }+ R_g~\phi \bigg]
-\int_\mathcal{M} d^2x \sqrt{g} g^{ab} \partial_a A~ \partial_b A  \\ & ~~~~-\int_\mathcal{M} d^2x \sqrt{g} R_g A - 2\int_\mathcal{M} d^2x \partial_a [\sqrt{g}g^{ab} (\phi-A)\partial_b A]
\end{split}
\ee
and note that the last term on the RHS above is a total derivative, and contribute to the boundary term. Also, it can be further checked that
\be  \label{term2}
\begin{split}
2 \int_{\partial \mathcal{M}}ds \sqrt{\tilde{h}}~ \tilde{K}_{\tilde{g}}~  \tilde{\phi} =& 2 \int_{\partial \mathcal{M}}ds \sqrt{h}~ K_g~\phi-2\int_{\partial \mathcal{M}}ds \sqrt{h}~ K_g~A \\
&+2 \int_{\partial \mathcal{M}}ds \sqrt{h}(\phi-A)n^a \partial_a A
\end{split}
\ee
The last term on the RHS of  \eqref{term2} and the last term, i.e. the boundary term, on the RHS of \eqref{term1} will cancel each other. Therefore, we can combine  \eqref{term1} and  \eqref{term2} to obtain
\be \label{tildeSL}
\begin{split}
 S_L[\tilde{\phi},\tilde{g}_{ab}]  = &S_L[\phi,g_{ab}] - \frac{c}{24\pi}\int_\mathcal{M} d^2x \sqrt{g} \bigg[g^{ab} \partial_a A \partial_b A +R_g A \bigg] \\
 &-\frac{c}{12\pi} \int_{\partial \mathcal{M}}ds \sqrt{h} K_g A.
 \end{split}
\ee
Note that the extra terms involving $A$, the second third term, on the RHS of \eqref{tildeSL} looks similar to Liouville action for the field $A$ except for the missing potential $e^{2A}$ term. This motivates us to add an extra term : $-\int_\mathcal{M} d^2x \sqrt{g}$ in the Liouville action $S_L$ in \eqref{defSL}, such that we now define an ``improved Liouville action'' $I_L[g_1,g_2]$
($g_1$ is the final metric and $g_2$ is the reference metric) as follows
\be \label{defSLa}
\begin{split}
I_L[e^{2\phi}g,g] =~ &\frac{c}{24\pi}\int_\mathcal{M} d^2x \sqrt{g} \bigg[g^{ab} \partial_a \phi \partial_b \phi+(e^{2\phi}-1)+R_g \phi \bigg] \\
&+\frac{c}{12\pi} \int_{\partial \mathcal{M}}ds \sqrt{h} K_g \phi.
\end{split}
\ee
For this action, we can find the following relation:
\be
I_L[e^{2\ti{\phi}}\ti{g},\tilde{g}] = I_L[e^{2\phi}g,g] - I_L[\ti{g},g].
\ee
This expression is naturally interpreted that the action $I_L[e^{2\ti{\phi}}\ti{g},\tilde{g}]$ measures the difference between the
final metric $e^{2\ti{\phi}}\ti{g}=e^{2\phi}g$ and the reference metric $\ti{g}=e^{2A}g$. In other words, this relation shows the chain rule:
\be
I_L[g_1,g_2]+I_L[g_2,g_3]=I_L[g_1,g_3], \label{chainr}
\ee
which includes the identity $I_L[g_2,g_1]=-I_L[g_1,g_2]$. The different between $S_L$ and $I_L$ does not depend on the Liouville field $\phi$ and thus the equations of motion from the variations of $\phi$ in both actions are the same. It is useful to note that if $g_1$ is the flat metric and the transformation from $g_1$ and $g_2$ is holomorphic, then we have $I_L[g_2,g_1]=0$, in other words the second plus third term in (\ref{tildeSL}) is vanishing.

In summary, $S_L[\phi,g]$ does not provide us with an absolute quantity which measures the complexity of the optimized state because it depends not only on the final metric $e^{2\phi}g$ but also on the reference metric $g$. Rather, we find it is convenient to look at the relative quantity $I_L[g_2,g_1]$ which is expected to measure the difference of complexity between the path-integral in $g_2$ and $g_1$.

 Before we move onto explicit evaluations of $I_L[g_2,g_1]$, we would like to mention that
 another potential source of ambiguity. We need to be careful with the fact that a constant shift of $\phi$ can change the action $S_L$ when the Euler number is non-zero due to the Gauss-Bonnet term $\int R_g\phi+2\int K_g\phi$ in $S_L$. This is removed by placing the background charge
as in the standard computation of correlation in Liouville CFTs \cite{Zam} and we will follow this prescription.

\subsection{Vacuum States}
Let us start with vacuum states in 2D CFTs on a circle with the circumference $2\pi$.
In AdS$_3/$CFT$_2$, they are dual to the global AdS$_3$. As we explained in section \ref{opt:pr}, we obtained the Poincare disk metric (\ref{pdiskm}) after the optimization. This metric can be written in the following two ways:
\ba
&& ds^2=e^{2\phi}(dr^2+r^2d\theta^2),\ \ \  e^{2\phi}=\frac{4}{(1-r^2)^2}, \label{metoone} \\
&& ds^2=e^{2\ti{\phi}}(dz^2+d\theta^2),\ \ \ e^{2\ti{\phi}}=\frac{1}{\sinh^2 z}, \label{metotwo}
\ea
where $\theta$ has a periodicity $2\pi$. We introduce the cut off $z=\ep$ and $r=r_0$ such that
$\frac{2r_0}{1-r_0^2}=\frac{1}{\ep}$ or equally $r_0\simeq 1-\ep+\frac{\ep^2}{2}+...$.
We express the flat metric for the polar coordinate $(r,\theta)$ and the Cartesian one $(z,\theta)$ by $g_{(r,\theta)}$ and $g_{(z,\theta)}$, respectively. Also the Poincare disk metric
$e^{2\phi}g_{(r,\theta)}=e^{2\ti{\phi}}g_{(z,\theta)}$ is represented by $g_{AdS}$.\\
The Liouville action for (\ref{metoone}) is evaluated as
\be \label{optijwidww}
\begin{split}
S_L[\phi,g_{(r,\theta)}] &= \frac{c}{24\pi}\int^{r_0}_0 rdr \int^{2\pi}_0 d\theta\left[(\de_r\phi)^2+e^{2\phi}\right]
+\frac{c}{12\pi}\int ds K_0\phi \\
&= \frac{c}{12}\int^{r_0}_0 dr\left(\frac{4(r^3+r)}{(1-r^2)^2}\right)+\frac{c}{6}\left(\phi(r=r_0)-\phi(r=0)\right)\\
&=\frac{c}{12}\left(\frac{2}{\ep}+2\log \ep-2+2\log 2\right)+\frac{c}{6}(-\log\ep-\log2) \\
&=\frac{c}{6}\left(\frac{1}{\ep}-1\right).
\end{split}
\ee
In the above, the contribution $\propto -\phi(r=0)=-\log 2$ comes from the background charge, while another one $\propto \phi(r=r_0)=-\log \ep$ is the standard boundary contribution. Finally, as before, $K_0$ corresponds the the trace of the extrinsic curvature of the boundary evaluated in the flat metric.

Similarly we can evaluate the Liouville action for (\ref{metotwo})
\be
S_L[\ti{\phi},g_{(z,\theta)}] =\frac{c}{12} \int^{z_\infty}_\ep\left[\frac{\cosh^2 z+1}{\sinh^2 z}\right] = \frac{c}{6}\left(\frac{1}{\ep}-1+\frac{z_{\infty}}{2}\right),  \label{gatk}
\ee
where $z_\infty(\to \infty)$ is the IR cut off in the $z$ integral. Indeed this expression differs from (\ref{optijwidww}).

We can also calculate the Liouville action for the Weyl scaling $g_{(r,\theta)}=e^{-2z}g_{(z,\theta)}$
\be
S_L[-z,g_{(z,\theta)}]=\frac{c}{12}\cdot \int^{z_\infty}_\ep(1+e^{-2z})= \frac{c}{24\pi}\left(\pi+2\pi z_{\infty}\right).  \label{gaqwwtk}
\ee

In terms of the improved Liouville action, we can summarize our results as follows:
\ba
&& I_L[g_{AdS},g_{(r,\theta)}]=\frac{c}{6\ep}-\frac{5c}{24}, \label{eewera} \\
&& I_L[g_{AdS},g_{(z,\theta)}]=\frac{c}{6\ep}-\frac{c}{6}, \label{eewerb} \\
&& I_L[g_{(r,\theta)},g_{(z,\theta)}]=\frac{c}{24}, \label{eewerc}
\ea
which indeed satisfy (\ref{chainr}).

\subsection{Primary States}

As we have seen in section (\ref{opt:pr}), for the primary states, the optimized metric is given by the conical geometry (denoted by $g_C$):
\ba
&& ds^2=e^{2\phi}(dr^2+ r^2 d\theta^2),\ \ \ e^{2\phi}=\frac{4 a^2}
{r^{2(1-a)} (1-r^{2a})^2}, \label{cgedo} \\
&& ds^2=e^{2\ti{\phi}}(dz^2+d\theta^2),\ \ \ e^{2\ti{\phi}}=\frac{a^2}
{\sinh^2\left(a z \right)}, \label{cgeo}
\ea
where we define $r=e^{-z}$ with $-z_{\infty}<z<-\epsilon$, $\delta(\equiv e^{-z_\infty})<r<r_0$ and $0\leq \theta<2\pi$. The UV cut off $r=r_0$ is specified as
\be
\frac{2 a r_0^{a}}{1-r_0^{2a}}=\frac{1}{\ep},
\ee
which is solved as $r_0^a=-a\ep+\s{1+(a \ep)^2}\simeq 1-a \ep+ (a \ep)^2/2+\ddd$.\\
The Liouville action for (\ref{cgeo}) is evaluated as follows:
\be
\begin{split}
S_L[\ti{\phi},g_{(z,\theta)}]=\frac{ca^2}{12}\int^{z_\infty}_\ep dz \left[\frac{\cosh^2 (a z)+1}{\sinh^2 (a z)}\right]
=\frac{c}{24\pi}\left(\frac{4\pi}{\ep}-4\pi a +2\pi a^2 z_{\infty}\right).
\end{split} \label{catk}
\ee
On the other hand, the Liouville action for (\ref{cgedo}) becomes
\be
\begin{split}
S_L[\phi, &g_{(r,\theta)}] = \frac{c}{12}\int^{r_0}_{\delta} rdr[(\de_r\phi)^2+e^{2\phi}]
+4\pi(\phi(r_0)-\phi(\delta)) \\
=& \frac{c}{12}\int^{z_\infty}_{\ep} dz \left[(\de_z\ti{\phi}+1)^2+e^{2\ti{\phi}}\right]
+ \frac{c}{6}\log \frac{\delta}{r_0}+\frac{c}{6}(\ti{\phi}(\ep)-\ti{\phi}(z_\infty)) \\
=& S_L[\ti{\phi},g_{(z,\theta)}]-\frac{c}{12} z_{\infty}.
\end{split}
\ee
In summary, we obtain the relative actions:
\be
\begin{split}
& I_L[g_C,g_{(r,\theta)}]=\frac{c}{6\ep}-\frac{ca}{6}-\frac{c}{24}+\frac{c}{12}(a^2-1)z_{\infty}, \\
& I_L[g_C,g_{(z,\theta)}]=\frac{c}{6\ep}-\frac{ca}{6}+\frac{c}{12}(a^2-1)z_{\infty},
\\
& I_L[g_{(r,\theta)},g_{(z,\theta)}]=\frac{c}{24},
\end{split}\label{eewerdd}
\ee
which indeed satisfy (\ref{chainr}). When $a=1$, they are reduced to (\ref{eewera}), (\ref{eewerb}), (\ref{eewerc}). For $a\neq 1$, the actions for $g_C$ include IR divergences, which may be canceled by further adding a background charge at the conical singularity.

\subsection{Finite Temperature State}

Finally we turn to the thermofield double (TFD) states dual to BTZ black holes. As discussed in section \ref{opt:ft},
after the optimization we obtained the metric of Einstein-Rosen bridge (denoted by $g_{ER}$):
\ba
&& ds^2=e^{2\phi}(dz^2+d\theta^2),\ \ \ e^{2\phi}=\frac{\left(\frac{2\pi}{\beta}\right)^2}
{\cos^2\left(\frac{2\pi z}{\beta}\right)}, \label{btzwq} \\
&& ds^2=e^{2\ti{\phi}}(dr^2+r^2d\theta^2),\ \ \ \ti{\phi}=\phi-\log r, \label{btqzwq}
\ea
where $r=e^{-z}$ with $-\beta/4<z<\beta/4$ and $0\leq\theta<2\pi$.\\
For the metric (\ref{btzwq}), the action is evaluated as follows:
\be
\begin{split}
S_L[\phi,g_{(z,\theta)}]&=\frac{c}{12}\int^{\beta/4-\ep}_{-\beta/4+\ep}dz
\left[\frac{4\pi^2}{\beta^2}\left(\tan^2(2\pi z/\beta)+\cos^{-2}(2\pi z/\beta)\right)\right] \\
&=\frac{c}{6}\left(\frac{2\pi}{\beta}\right)^2\cdot \int^{\beta/4-\ep}_0
dz \left[\frac{1+\sin^2(2\pi z/\beta)}{\cos^2(2\pi z/\beta)}\right] \\
&= \frac{c}{3\ep}-\frac{\pi^2c}{6\beta}.
\end{split}\label{btpp}
\ee
On the other hand for the metric (\ref{btqzwq}), the action is evaluated as follows:
\be
\begin{split}
S_L[\ti{\phi},&g_{(r,\theta)}]=\frac{c}{12}\int^{r_2}_{r_1} rdr\left[(\de_r\ti{\phi})^2+e^{2\ti{\phi}}\right]
+\frac{c}{6}(\ti{\phi}(r=r_2)-\ti{\phi}(r=r_1))\\
&=\frac{c}{12} \int^{\beta/4-\ep}_{-\beta/4+\ep} dz ((\de_z\phi-1)^2+e^{2\phi})
+\frac{c}{6}(\ti{\phi}(r=r_2)-\ti{\phi}(r=r_1))\\
&=\frac{c}{3\ep}-\frac{\pi^2 c}{6\beta}-\frac{c}{24}\beta,
\end{split}  \label{btzfind}
\ee
where $r_1=e^{-\frac{\beta}{4}+\ep}$ and $r_2=e^{\frac{\beta}{4}-\ep}$.\\
The relative action is computed as follows
\ba
&& I_L[g_{ER},g_{(z,\theta)}]=\frac{c}{3\ep}-\frac{\pi^2c}{6\beta}-\frac{c}{24}\beta, \label{eewbtzr}\\
&& I_L[g_{ER},g_{(r,\theta)}]=\frac{c}{3\ep}-\frac{\pi^2 c}{6\beta}
-\frac{c}{24}\beta-\frac{c}{24}(e^{\beta/2}-e^{-\beta/2}), \label{eewbtzrs} \\
&& I_L[g_{(r,\theta)},g_{(z,\theta)}]=\frac{c}{24}(e^{\beta/2}-e^{-\beta/2}),
\ea
which again satisfy (\ref{chainr}).

\section{Application to NAdS$_2/$CFT$_1$}\label{AdS2CFT1}

As recently discovered, to make sense of AdS$_2/$CFT$_1$, we need a conformal symmetry breaking term \cite{SYK,Dilaton,Dilatona,Dilatonb}, under the reparameterization of
$\ti{\tau}=\ti{\tau}(\tau)$, often called NAdS$_2/$CFT$_1$ duality. The effective action is written as a Schwarzian derivative term $Sch[\ti{\tau},\tau]$ as explicitly realized in the Sachdev-Ye-Kitaev (SYK) model \cite{SY,Kit}:
\be
Sch[\ti{\tau},\tau]=-\frac{3}{2}
\left(\frac{\de^2_\tau\ti{\tau}}
{\de_\tau\ti{\tau}}\right)^2+\frac{\de^3_\tau\ti{\tau}}{\de_\tau\ti{\tau}}. \label{schacthe}
\ee

For the one dimensional metric $ds^2=e^{2\phi}d\tau^2$, we can identify
 $\frac{d\ti{\tau}}{d\tau}=e^{\phi}$. Thus the conformal symmetry breaking term (\ref{schacthe}) looks like $N\int d\tau (\de_\tau\phi)^2$, where $N$ is a constant proportional to degrees of freedom. Therefore we find (we shifted $\phi$ appropriately)
\ba
&& \Psi_{g_{\tau\tau}=e^{2\phi}}(\ti{\vp}(x))=e^{S_{1}[\phi]-S_1[0]}\cdot \Psi_{g_{\tau\tau}=1}(\ti{\vp}(x)),  \no
&& S_{1}[\phi]=N\int d\tau\left[(\de_\tau\phi)^2+2e^{\phi}\right].
\ea
By minimizing the action, this again leads to
\be
ds^2=e^{2\phi}d\tau^2=\frac{d\tau^2}{\tau^2}.
\ee
This is consistent with the time slice of AdS$_2$ space-time. Note that if there is no conformal symmetry breaking effect, we cannot stabilize the optimization procedure. Also notice that in standard tensor network descriptions, it is very difficult to describe one dimensional quantum mechanics as we normally coarse-grain space directions to build an extra dimension  in the network. In our path-integral approach the extra dimension arises naturally even in quantum mechanics.

\section{Applications to Higher Dimensional CFTs}\label{HIGHERDIM}

Higher dimensional generalizations of our optimization procedure do not seem to be
straightforward as the generic metric cannot be expressed only by
the Weyl scaling as in (\ref{met}). Nevertheless, it is useful to see
what optimization can lead to correct time slices of gravity duals by taking into account only the Weyl scaling degrees of freedom as a first step toward this direction. As we will see below, at least for pure AdS$_{d+1}$ we can obtain expected results even from this limited range of optimization.

\subsection{Our Formulation}
For this we need a complexity functional $I[\phi]$ for the metric of the form:
\be
ds^2=e^{2\phi(x)}g_{ab}dx^adx^b, \label{hmetr}
\ee
with $x$ regarded as $d$ dimensional vector (which includes ``$z$'' coordinate).
 We propose that for a vacuum state in a $d$ dimensional CFT, the optimization can be done by minimizing the following functional $I^{bulk}_d[\phi,g]$ ($N$ is a normalization factor proportional to
 the degrees of freedom):
\be
I^{bulk}_d[\phi,g]=N\int_{\Sigma} dx^{d} \s{g} \left[e^{d\phi}+e^{(d-2)\phi}\left(g^{ab}\de_a\phi\de_b\phi\right)
+\frac{e^{(d-2)\phi}}{(d-1)(d-2)}R_g\right],\label{hacd}
\ee
where $R_g$ is the Ricci scalar for the reference metric $g$. Reader can regard this as a generalization of Liouville action to general dimensions, which is quadratic in derivative of $\phi$ field\footnote{The possibility of having terms in the complexity functional with higher than quadratic derivatives of $\phi$ is discussed in section \ref{hdanom} and in appendix \ref{SZcase}. In fact they will be important for reproducing correct anomalies for even dimensional CFTs in higher dimensions.}. The computation of entanglement entropy shown later allows us to identify the normalization factor $N$ in (\ref{hacd}) for holographic CFTs:
\be
N=\frac{(d-1)R^{d-1}}{16\pi G_N}, \label{normnnn}
\ee
where $R$ is the AdS radius In particular for $d=2$ and $d=4$ we find
\ba
&& N_{d=2}=\frac{c}{24\pi},\no
&& N_{d=4}=\frac{3}{2\pi^2}a_4, \label{normnnnn}
\ea
in terms of the central charge $c$ in 2D CFTs and $a_4$ in 4D CFTs.

Indeed, the minimization of $I^{bdy}_d[\phi,g]$ leads to the hyperbolic space H$_d$ which is the time slice of pure AdS$_{d+1}$ as we will see later. For the optimization of reduced density matrix we need to introduce the boundary term as in section \ref{HEE2DCFT}.
We argue it is given by
\be
I^{bdy}_d[\phi,g]=2N\int_{\de\Sigma} dx^{d-1} \s{\gamma}\left[\frac{K_g}{d-1}\frac{e^{(d-2)\phi}}{d-2}+\mu_B \frac{e^{(d-1)\phi}}{d-1}\right],\label{hdbd}
\ee
where $\gamma_{ij}$ is the induced metric on the boundary $\de\Sigma$.
This again leads to the boundary condition $K+(d-1)\mu_B=0$, where $K=e^{-\phi}((d-1)\de_n\phi+K_0)$.
We defined our optimization by minimizing the total functional $I^{tot}_d[\phi,g]$
\be
I^{tot}_d[\phi,g]=I^{bulk}_d[\phi,g]+I^{bdy}_d[\phi,g].  \label{totocompw}
\ee

It is important to consider the limit $d \rightarrow 2$ in $I^{tot}_d[\phi,g]$ and explore the possibility of recovering the standard Liouville action for $d=2$ dimensions. As it is obvious from the expression of $I^{bulk}_d[\phi,g]$ in (\ref{hacd}), a naive limit of $d\rightarrow 2$ is singular as the third term on the RHS of (\ref{hacd}) gives a contribution proportional to $1/(d-2)$. However, as it was mentioned in section \ref{weyl2d} (see the discussion in the paragraphs following (\ref{defSLa})), for an absolute measure of complexity in 2-dimensions the relative or the improved Liouville action $I_L(g_1,g_2)$, defined in  (\ref{defSLa}), is more suitable compared to $S_L$, defined in  (\ref{defSL}). The advantage was mentioned to be the fact that $I_L(g_1,g_2)$ being a relative  measure of complexity does not depend on the reference metric. It is interesting to note that the subtlety of taking the $d\rightarrow 2$ limit works out perfectly if we subtract away the contribution of the reference metric while taking the above mentioned limit. Therefore we notice the following identity
\be
\lim_{d \to 2}\bigg[I^{tot}_d[\phi,g]-I^{tot}_d[\phi= 0,g]\bigg] =I_L[e^{2\phi}g,g],
\ee
where $g_{ab}$ is considered as the reference $2$-dimensional metric and therefore following the discussion after  (\ref{defSLa}), $I_L[e^{2\phi}g,g]$ computes the relative complexity of the generic metric $e^{2\phi} g_{ab}$ compared to that reference metric.

Notice that actually we can combine the functional $I_d+I^{bdy}_d$ into the Einstein-Hilbert action
plus a cosmological constant on the final metric (\ref{hmetr}) which we write $\ti{g}=e^{2\phi}g$
and $\ti{\gamma}=e^{2\phi}\gamma$:
\ba
I^{tot}_d&=&N\int_{\Sigma}dx^d \s{\ti{g}}\left[1+\frac{R_{\ti{g}}}{(d-1)(d-2)}\right]\no
&&+2N\int_{\de\Sigma} dx^{d-1}\s{\ti{\gamma}}\left[\frac{K_g}{(d-1)(d-2)}+\frac{\mu_B}{d-1}\right]. \label{EHopt}
\ea
From this manifestly covariant expression which only depends on $\ti{g}$, as opposed to the 2D Liouville action, the
invariance of the action by the change of reference metric is manifest:
\be
I^{tot}_d[\phi-A,e^{2A}g]=I^{tot}_d[\phi,g].
\ee

\subsection{$AdS_{d+1}$ from Optimization}

Here we would like to confirm that the optimization leads to expected AdS geometries for
vacuums states. This is almost obvious from the expression (\ref{EHopt}). However notice that we
 take only the Weyl mode $\phi$ dynamical.

Consider a CFT$_{d}$ defined on R$^d$ or R$\times$S$^{d-1}$. In these two cases the metrics are taken in the following form:
\ba
{\mbox R}^d &:&\ \ ds^2=e^{2\phi(z)}\left(dz^2+\sum_{i=1}^{d-1}dx_i^2\right), \label{aidwjbnd}\\
{\mbox R}\times {\mbox S}^{d-1} &:& \ \ ds^2=e^{2\phi(r)}(dr^2+r^2d\Omega_{d-1}^2). \label{aidwjbndd}
\ea
 The values of the functional $I_d$ in these cases read
 \ba
{\mbox R}^d &:&~I^{bulk}_d=N\int dx^d\left[e^{d\phi}+e^{(d-2)\phi}(\de_z\phi)^2\right], \label{idwsjbnd}\\
{\mbox R} \times {\mbox S}^{d-1} &:&~I^{bulk}_d=N\int d\Omega_{d-1}\int dr \cdot r^{d-1}\left[e^{d\phi}+e^{(d-2)\phi}(\de_r\phi)^2\right], \label{aaidwjbndd}
 \ea
 and their equation of motions are given by
 \ba
 {\mbox R}^d&:& ~d e^{2\phi}-(d-2)e^{(d-2)\phi}(\de_z\phi)^2-2\de_z^2\phi=0, \label{eomhftg}\\
{\mbox R} \times {\mbox S}^{d-1}&:&~d e^{2\phi}-(d-2)e^{(d-2)\phi}(\de_r\phi)^2-2\de_r^2\phi-\frac{2(d-1)}{r}\de_r\phi=0. \no \label{eomrhftg}
 \ea
We can confirm that both of them have the hyperbolic space solutions:
\ba
 {\mbox R}^d &:& ~e^{2\phi(z)}=\frac{1}{z^2}, \label{rdsolk} \\
 {\mbox R}\times {\mbox S}^{d-1} &:& ~ e^{2\phi(r)}=\frac{4}{(1-r^2)^2}. \label{rdsolkk}
\ea
They coincide with the time slice of AdS$_{d+1}$ as expected.

\subsection{Excitations in Global AdS$_{d+1}$}

Now let us consider excitations in a $d$ dimensional CFT on R$\times$S$^{d-1}$.
We focus on the case $d=3,4$ and assume that the excitations are homogeneous and static.
In AdS$_{d+1}/$CFT$_d$, such a state is dual to spherically symmetric solution
given by the AdS$_4$ Schwarzschild black hole solution
\be
ds^2=-h(\rho) dt^2+\frac{d\rho^2}{h(\rho)}+\rho^2d\Omega_{d-1}^2. \label{AdSbhqwe}
\ee
with $h(\rho) = \rho^2+1-M\rho^{-(d-2)}$.
Here we are interested in the leading correction when $M$ is very small.
We focus on the time slice $t=0$ and rewrite
\be
ds^2=\frac{d\rho^2}{h(\rho)}+\rho^2d\Omega_{d-1}^2=e^{2\phi}(dr^2+r^2d\Omega_{d-1}^2).
\label{funcphiy}
\ee
We can find an explicit relation between $\rho$ and $r$ and the function $\phi$ as follows
(up to the linear order of $M$)
\ba
&& r\simeq \frac{\rho}{1+\s{1+\rho^2}}\cdot (1+M f(\rho)),\no
&& e^{\phi(r)}\simeq (1+\s{1+\rho^2})\cdot (1-M f(\rho)),
\ea
where the function $f(r)$ depends on the dimension $d$
\ba
&& d=3:\ \ \ f(\rho)=1-\frac{2\rho^2+1}{2\rho\s{\rho^2+1}},\no
&& d=4:\ \ \ f(\rho)= -\frac{1}{4} \left[\frac{3 \rho ^2+1}{\rho ^2 \sqrt{\rho ^2+1}} +3 \log \left[{\rho \over\sqrt{\rho ^2+1}+1 }\right]\right].
\ea
Finally we obtain the function $\phi(r)$ in (\ref{funcphiy}) in the form
\be  \label{exgads5eq5}
e^{\phi(r)} \simeq  {2  \over (1-r^2)}  \bigg(1 + M \cdot \eta(r)\bigg),
\ee
where the function $\eta(r)$ is explicitly given for each $d$:
\ba
&& d=3:\ \ \eta(r)=\frac{(1-r)^3}{4r(1+r)},\label{wweqhjhj} \\
&& d=4:\ \ \eta(r) =\frac{r^6+9 r^4-9 r^2-12 \left(r^4+r^2\right) \log (r)-1}{16 r^2 \left(r^2-1\right)}.\label{wwesqhjhj}
\ea

On the other hand, the optimization of our functional (\ref{hacd}) given by the differential equation (\ref{eomrhftg}) leads to the perturbative solution of the form:
\be \label{pertdef5}
e^{\phi} = {2 \over (1-r^2)} \bigg( 1+ \ti{M}\cdot h(r) \bigg) +O(\ti{M}^2),
\ee
where we treat $\ti{M}$ as an infinitesimally small parameter. We can analytically determine the function $h(r)$ and confirm that $h(r)$ is equal to $\eta(r)$ in (\ref{wweqhjhj}) and
(\ref{wwesqhjhj}) in each dimension up to a constant factor.

In this way we find that the first order back-reaction to the time slice metric in AdS gravity is correctly reproduced by our optimization procedure.

\subsection{Holographic Entanglement Entropy}

In this subsection we will show that the total action $I^{bulk}_d+I^{bdy}_d$ can reproduce the correct holographic entanglement entropy (HEE) \cite{RT} when the subsystem $A$ is a round ball. We will also focus on $d=3,4$ case in the AdS$_{d+1}/$CFT$_{d}$.

We will closely follow the method that was explicitly used in the case of 2D CFT in section \ref{HEE2DCFT}. We start with the holographic construction of density matrix and argue that this accurately reproduce the entanglement wedge as expected and following that we will compute the entanglement entropy holographically.

The metric of the manifold on which the path-integral for the density matrix is being computed, will be denoted by
\be
ds^2 = e^{2\phi}\left(dz^2 + \sum_{i=1}^{d-1}dx_i^2 \right) =  e^{2\phi}\left(dz^2 + dr^2 +r^2 d\Omega_{d-2}^2 \right),
\ee
where, sometimes, we will also use the notation $ds^2 = e^{2\phi}g_{ab}dx^adx^b$, with the understanding that the reference metric $g_{ab}$ is the flat metric $g_{ab}dx^adx^b = dz^2 + \sum_{i=1}^{d-1}dx_i^2 = dz^2 + dr^2 +r^2 d\Omega_{d-2}^2$. Also,  $d\Omega_{d-2}^2$ is the metric for $(d-2)$-dimensional unit sphere. Therefore, we will have
\be
\begin{split}
d=3 ~~ \Rightarrow ~~& d\Omega_{1}^2 = d\theta^2 \\
d=4 ~~ \Rightarrow ~~& d\Omega_{2}^2 = d\theta^2 + \sin^2\theta ~d\phi_1^2
\end{split}
\ee
The round ball subsystem is defined by $A_D = \{x_i | r\le \ell \}, ~r=\sqrt{\sum_{i=1}^{d-1}x_i^2}$, where $\ell$ is the radius of the circular disk.

Following the same steps, as depicted in section \ref{HEE2DCFT}, we should proceed with the optimization that will lead us to identifying the boundaries $\Gamma^{(d)}_{A}$ of the bulk regions $\Sigma^{(d)}_{\pm}$. The boundary condition for $\phi$ should be imposed at the two edges of the slit composed by the boundary of the round ball $A_D$, i.e. at $r=\ell$ near $z=0$. These arguments validate that we should also consider the boundary part of the action $I^{bdy}_d$
\be \label{Sbd}
\begin{split}
I^{bdy}_d=&2N\int_{\de\Sigma} d^{d-1}x \s{\gamma}\left[\frac{K_g}{d-1}\frac{e^{(d-2)\phi}}{d-2}+\mu_B \frac{e^{(d-1)\phi}}{d-1}\right]
\end{split}
\ee
Let us describe the boundary $\Gamma_A$ as $r=f(z)$ with the normal vectors (normalized with respect to the full metric $e^{2\phi} g_{ab}$) given by
\be
n^z = -{e^{-\phi} f'(z) \over \sqrt{1+f'(z)^2}}, ~~n^r={e^{-\phi}  \over \sqrt{1+f'(z)^2}}, ~~n^{\Omega_{d-2}}=0.
\ee
The extrinsic curvature of the boundary is
\be
K = \gamma^{ab}\nabla_a n_b = e^{-\phi} (K_0 +(d-1)n^a\partial_a\phi).
\ee
where $K_0$ is the extrinsic curvature for the same boundary but in the reference metric $g_{ab}$ (which is flat metric in our case)
\be
K_0 =-\frac{f''}{(1+(f')^2)^{3/2}}.
\ee
The boundary condition for the field on the edge of the slit (the round ball $A_D$ in our case) is Dirichlet, however we need to impose Neumann boundary condition on the surface $\Gamma_A$ and it leads us to the condition
\be
K +(d-1) \mu_B=0.
\ee
For the determination of the density matrix, since the two boundaries $\Gamma_A$ in $\Sigma_+$
and that in $\Sigma_{-}$ are pasted smoothly, we should consider $\mu_B=0$. The optimization determines the bulk metric to be the hyperbolic one $\phi = -\log z$.

In order to fix the shape of $\Gamma_A$ we should solve $K=0$, which is precisely the condition of minimal surfaces. Thus we find that $\Gamma_A$ is given by the half-sphere $z^2+r^2 = \ell^2$. Accordingly, the holographic dual for the density matrix corresponds to the region $z^2+r^2 \le \ell^2$ and it agrees with the entanglement wedge.

Let us now consider the entanglement entropy and for that we need to consider $\rho^n_A$ and finally analytically continue considering $|n-1| \ll 1$, leading us to a conical geometry with deficit angle $2\pi(1-n)$ along the entangling surface $r=\ell$. It is then natural to expect that the extrinsic curvature for the boundaries $\Gamma_A$ will now become different from vanishing, i.e. $K \neq 0$. This can be estimated from considering that the boundaries $\Gamma_A$ now changes from $z^2+r^2 = \ell^2$ to $(z-z_0)^2+r^2 = \ell^2$, with which we can now evaluate the extrinsic curvature
\be
K= -(d-1) {z_0 \over l}.
\ee
Also with infinitesimal $z_0$ we obtain $r \sim l + {z_0 \over l}z + \mathcal{O}(z^2)$ near the boundary point $\{z,r,\Omega^{(d-2)}\} = \{0,\ell,\Omega^{(d-2)}\}$. The corner angle at the intersection of $\partial \Sigma^{(d-1)}_{\pm}$ and the entangling surface $r=\ell$, also becomes $ \pi / 2  + z_0 / l$. For the n-sheeted conical geometry we can interpret this corner angle as $ z_0 / l = \pi (1-n)$ and thus obtain the relation
\be
K= (d-1) (n-1).
\ee
We need to satisfy the Neumann boundary condition $K +(d-1) \mu_B=0$ at $\partial \Sigma^{(d-1)}_{\pm}$, and we implement this by setting
\be
\mu_B = \pi (1-n).
\ee
Note that this condition is true for any dimension $d$.

\paragraph{The entanglement entropy in $3$-d CFT :}
Now one can explicitly check that the holographic entanglement entropy can be evaluated for $d=3$ by considering $\mu_B = \pi (1-n)$ in the boundary action in  \eqref{Sbd} as follows
\be \label{heed3}
\begin{split}
S_A^{(d=3)} =& - \partial_n \left[2N \mu_B \int_{\partial \Sigma^{(2)}_{+}} {e^{2\phi} \over 2} + 2N \mu_B \int_{\partial \Sigma^{(2)}_{-}} {e^{2\phi} \over 2} \right] \\
= &4N\pi^2 \left[{\ell \over \epsilon} -1 \right]
\end{split}
\ee
where $\epsilon$ is the UV cut-off as the range of integration in $r$ has been taken to be $\epsilon \le r \le \ell$.
\paragraph{The entanglement entropy in $4$-d CFT :}
Similarly for $d=4$ one obtains
\be \label{heed4}
\begin{split}
S_A^{(d=4)} =& - \partial_n \left[2N \mu_B \int_{\partial \Sigma^{(3)}_{+}} {e^{3\phi} \over 3} + 2N \mu_B \int_{\partial \Sigma^{(3)}_{-}} {e^{3\phi} \over 3} \right] \\
= & { 8N\pi^2 \over 3}\left[{\ell^2 \over \epsilon^2} -\log\left({\ell \over \epsilon}\right) + \left({1\over 2} + \log 2\right)\right]
\end{split}
\ee

Finally, It can be checked that the expressions in both \eqref{heed3} and \eqref{heed4} do indeed reproduce the correct behavior for holographic entanglement entropy in higher dimensions, compare with \cite{RT}, by choosing the normalization $N$ as in (\ref{normnnn}) and (\ref{normnnnn}).\\
Moreover, we confirm that for the spherical choice of the region the general formula for entanglement entropy reads
\be
S^{d}_A=\f{4\pi N}{d-1}\int_{\Gamma_A}e^{(d-1)\phi}.
\ee

\subsection{Evaluation of Complexity Functional} \label{evholcomp}
As we argued for 2D CFTs (see discussions following \eqref{newe}), the Liouville action, $S_L$ when computed on-shell for the solutions gives us a measure of holographic computational complexity. Here we would like to examine an analogous quantity for the higher dimensional CFTs. Namely, we evaluate the complexity functional $I^{bulk}_d+I^{bdy}_{d}$ for optimized solutions corresponding to the global AdS$_{d+1}$. We focus on the $d=3,4$ case again.
\paragraph{3D CFT (d=3) :}
Consider the metric obtained by setting $d=3$ in the solution of (\ref{rdsolkk}).
The boundary condition for $r=r_0$ is chosen as in 2D case:
\be \label{bdcond1}
{4r_0^2 \over (1-r_0^2)^2} = {1 \over \epsilon^2}~~ \Rightarrow ~~ r_0 = 1-\epsilon + {\epsilon^2 \over 2} + \mathcal{O} (\epsilon^3)
\ee
Then the bulk $I^{bulk}_d$ and boundary $I^{bdy}_d$ are evaluated as follows:
\be
\begin{split}
I^{bulk}_3=&4\pi N \left[{1 \over \epsilon^2} - {2 \over \epsilon} + {1 \over 2} + \log \left({2 \over \epsilon}\right)\right], \\
I_3^{bdy}=& 8 \pi N \left[{1  \over \epsilon} + {\mu_B \over 2\epsilon^2}\right].
\end{split}
\ee
Finally, adding the two contributions we obtain
\be \label{CFd3}
I_3^{tot} = 4\pi N \left[{1 \over \epsilon^2} + {1 \over 2} + \log \left({2 \over \epsilon}\right)\right] + {4 \pi N \mu_B \over \epsilon^2}.
\ee
\paragraph{4D CFT (d=4) :}
In the same way, when $d=4$, the solution (\ref{rdsolkk}) leads to
\be
\begin{split}
I^{bulk}_4=&2\pi^2 N \left[{2 \over 3\epsilon^3} - {1 \over \epsilon^2} +{1 \over \epsilon} -{5 \over 12}\right], \\
I_4^{bdy}=& 4 \pi^2 N \left[{1  \over 2\epsilon^2} + \mu_B \left({1 \over 3 \epsilon^3} + {1 \over 8} \right)\right].
\end{split}
\ee
Totally we obtain
\be \label{CFd4}
 I_4^{tot} = 2\pi^2 N \left[{2 \over 3\epsilon^3} +{1 \over \epsilon} -{5 \over 12} \right]+ 4 \pi^2 N \mu_B \left({1 \over 3 \epsilon^3} + {1 \over 8}  \right).
\ee

For the sake of comparing our results with the existing literature, which we do in the next sub-section, we have to set $\mu_B=0$ and the normalization factor $N$ in the above formulas should be taken as defined in (\ref{normnnn}) for holographic CFTs. Therefore, by simply setting $R_{AdS}=1$, the complexity $C_{\Psi_0}$ of the vacuum state $|\Psi_0\lb$ computed from our complexity functional $I^{tot}_d$, is given as follows
\ba
&& \mbox{3D CFT:}\ \ \ \  C^{3}_{\Psi_0}=\frac{1}{2G_N}\left[{1 \over \epsilon^2} + {1 \over 2} + \log \left({2 \over \epsilon}\right)\right] \label{tbbb} \\
&& \mbox{4D CFT:}\ \ \ \  C^{4}_{\Psi_0}=\frac{\pi}{8G_N} \left[{2 \over \epsilon^3} +{3 \over \epsilon} -{5 \over 4} \right] \label{tccc}.
\ea
It is also helpful to remember that for 2D CFTs, according to (\ref{eewerb}), we find the complexity of the vacuum (if we choose $g_{(z,\theta)}$ as the reference metric)
\be
 \mbox{2D CFT:}\ \ \ \  C^{2}_{\Psi_0}=I_L[g_{AdS},g_{(z,\theta)}]=\frac{1}{4G_N}\left(\frac{1}{\ep}-1\right).
 \label{taaa}
\ee
Notice that generally the complexity behaves like $C^{d}_{\Psi}\sim \ep^{-(d-1)}$
in any CFT$_{d+1}$ and this is interpreted as the volume law divergence.

\subsection{Comparison with Earlier Conjectures}

It is worthwhile to compare our complexity functional $I^{tot}_d$ evaluated specifically for dimensions $d=3,4$ in the previous subsection, with those in earlier conjectures.\\

{\bf (1) ``Complexity = Volume'' Conjecture}

Recently there has been exciting developments in understanding computational complexity in quantum systems holographically, i.e. some geometric calculation in the gravity side has been proposed to be measuring the computational complexity. The computational complexity of any boundary state at any given time, i.e. on some spacelike slice, say $\Sigma$, of the boundary, was first proposed in \cite{SUR,Susskind} to be identified with the volume of a maximal volume space-like slice, say $M_\Sigma$, in the bulk where the bulk space like surface ends on the given boundary slice. We will refer to it as the CV-conjecture (complexity = volume),
\be
C_V(\Sigma) ={\mathcal{V}(M_{\Sigma}) \over G_N R_{AdS}},~~\text{such that} ~~ \partial{M_{\Sigma}} = \Sigma  \label{CVcon}
\ee
where $\mathcal{V}(M_{\Sigma})$ denotes the volume of the maximal time slice $M_{\Sigma}$, and $R_{AdS}$ is some associated length scale in the bulk, conveniently taken to be the AdS radius in AdS/CFT.
For TFD states in holographic CFTs, the gravity dual computation of $C_V$ shows the linear growth
\cite{SUR,Susskind}.

It may also be useful to mentioned that in the paper \cite{InfoM}, it was found that
the gravity dual of the information metric $G(\Sigma)$ in CFT$_d$, which is equivalent to an integral of two point functions of a marginal primary operator, is well approximated by
$G(\Sigma)\simeq n_d\cdot {\mathcal{V}(M_{\Sigma}) \over R^{d}_{AdS}}$ ($n_d$ is a numerical constant).
Indeed, the information metric in CFTs also have the linear growth under the time evolution of
a TFD state.\\

{\bf (2) ``Complexity = Action'' Conjecture }

Later, in \cite{BrownSusskind1,BrownSusskind2}, it was also conjectured that the complexity is given by the action of a Wheeler-de Witt patch in the bulk bounded by the given space-like surface. One motivation for this conjecture was to remove the unpleasant feature about the CV-conjecture that it depends on a the choice of a length scale $R_{AdS}$. We will refer to it as the CA-conjecture (complexity = Action)
\be
C_A(\Sigma) = {I_{WDW} \over \pi  } \label{CAcon}
\ee
where $I_{WDW}$ is given by the Einstein-Hilbert action integrated only over the Wheeler-DeWitt (WDW) patch $M_{WDW}$, which extends from the boundary time slice $\Sigma$ where we measure the complexity. The WDW patch is defined to be the bulk space-time region in the bulk which is union of all the space-like surfaces anchored at the boundary at a given time of the CFT. It is easy to visualize this for the eternal black-hole Penrose diagram (see figure 1 in \cite{BrownSusskind1}). In that case, once we pick two given times at the two boundary CFTs, say $t_L$ and $t_R$ respectively, the WDW patch is the bulk space-time region bounded by the null surfaces and such that it is union of all possible space-like surfaces anchored at the times $t_L$ and $t_R$ in the boundary.

Schematically, $I_{WDW}$ can be written as
\be
I_{WDW} = {1 \over 16 \pi G_N} \int_{M_{WDW}} d^{d+1}x \sqrt{-g} (R-2\Lambda) + I^{bdy}_{WDW},
\label{actionBS}
\ee
where $I^{bdy}_{WDW}$ contains the important boundary contributions coming from the null boundaries of the WDW patch $M_{WDW}$, also including the joint contributions coming from the intersections of the null boundaries \cite{Lehner:2016vdi}.

In a related direction a quantity called complexity of formation was defined and studied in \cite{Chapman:2016hwi}. This quantity computes, following the CA conjecture, the difference of the action (only the bulk part) between the BTZ and two times that of the AdS (vacuum). The similar results were reproduced in \cite{Kim:2017lrw} with a proposal of renormalized holographic complexity.

It is important to mention that, through our proposal we can only learn about the fixed time behavior of the computational complexity for the dual state in the CFT, whereas, the original motivation of proposing a holographic measure of complexity was to study it's growth with time \cite{BrownSusskind1,BrownSusskind2,Lehner:2016vdi}. In order to compare the evaluation of complexity with our proposal with the same evaluated using other proposals, we need to therefore look into their constant time evaluations.
In \cite{CMR,Reynolds:2016rvl} the authors investigated the constant time behavior of the holographic complexity. More specifically they studied the divergence structure, considering both the CV and CA-conjectures. Also a possible prescription to remove an ambiguity due to different parametrization of the null boundary surfaces in the WDW patch was found in \cite{Lehner:2016vdi}. This prescription was used to evaluate the holographic complexity in \cite{Reynolds:2016rvl}. In appendix \ref{holcomlit}, we summarize these results of holographic complexity by focusing on the vacuum states. \\

{\bf Comparisons with Our Results }

We are finally ready to compare the evaluation of holographic complexity with our proposal against the same computed with the existing proposals in the literature, presented in appendix \ref{holcomlit}.

First if we follow the ``Complexity = Volume'' conjecture (\ref{CVcon}), the complexity
has the structure $C_V\sim c^{(1)}_v\cdot \ep^{-(d-1)}+c^{(3)}_v\cdot  \ep^{-(d-3)}+\ddd$.
This behavior agrees with our results of complexity $C_{\Psi_0}$ presented in (\ref{taaa}), (\ref{tbbb}) and (\ref{tccc}), though the
relative coefficients do not coincide in general.

Next we turn to the ``Complexity = Action'' conjecture (\ref{CAcon}). The analysis in \cite{CMR} evaluates it to be divergent, in fact a logarithmically enhanced divergence of the form $\log\ep^{-1}\cdot \epsilon^{(d-1)}$ for the CA-conjecture as opposed to the $1/\epsilon^{d-1}$ divergence for the CV-conjecture. On the other side, the \cite{Reynolds:2016rvl} proposal, which introduces an additional boundary contribution, produces a surprising result for the $d=2$ case i.e. bulk $AdS_3$: for both Poincare and global $AdS_3$, the leading divergence vanishes, leading to a constant holographic complexity. In higher dimensions $d=3,4$, the holographic complexity has a leading divergence of the form $1/\epsilon^{d-1}$ for both Poincare and global $AdS_{d+1}$.
 Therefore the divergence structure in \cite{Reynolds:2016rvl} for $d>2$ is the same as ours, whereas, they differ in the numerical coefficients in general. Nevertheless, in the next subsection, we will point out an interesting relation between our complexity functional $I^{tot}_d$ and the gravity action $I_{WDW}$ in the WDW patch.

Since there is no precise definition of computational complexity in quantum field theories known at present, we cannot decide which of these prescriptions is true by consulting with rigorous results in field theory. However, notice that our proposal of computational complexity $C_{\Psi_0}$, defined in (\ref{cpxidef}), is based on not any holography but a purely field theoretic argument as is clear in two dimensional CFT case, where it is related to the normalization of wave functional.

\subsection{Relation to ``Complexity = Action'' Proposal}
We have discussed in the previous subsection that, in \cite{BrownSusskind1,BrownSusskind2}, it has been conjectured that the holographic complexity is measured by the bulk action being integrated over the WDW patch defined above including suitable boundary terms. Here we would like to compare this quantity with our complexity functional. For simplicity, we set $R_{AdS}=1$ and thus  $\Lambda=-\frac{d(d-1)}{2}$ below.

Consider the following class of $d+1$ dimensional space-time:
\be
ds^2=-dt^2+\cos^2 t\cdot  e^{2\phi(x)}h_{ij}dx^idx^j, \label{wdwads}
\ee
where $t$ takes the range $-\pi/2\leq t\leq \pi/2$ and $i,j=1,2,\ddd,d$.
The pure AdS$_{d+1}$ which is a solution to the Einstein equation from $I_{WDW}$, is obtained when
the metric $e^{2\phi(x)}h_{ij}dx^i dx^j$ coincides with a hyperbolic space $H_{d}$.
For example, when $d=2$, the Einstein equation just leads to $(\de_1^2+\de_2^2)\phi=e^{2\phi}$ i.e. the Liouville equation. Note that in this pure AdS$_{d+1}$ solution, the coordinate covered by
(\ref{wdwads}) indeed represents the WDW patch.\footnote{In Euclidean signature obtained by
 $t\to i\tau$, this leads to the hyperbolic slice of H$_{d+1}$ which is precisely given by (\ref{ourm}).}
Motivated by this we identify this space (\ref{wdwads}) with $M_{WDW}$. However, note that for generic choices of $\phi$ and $h_{ij}$ (\ref{wdwads}) does not represent the WDW patch in the
original definition in \cite{BrownSusskind1,BrownSusskind2}. They coincide only on-shell.

Now we would like to evaluate the gravity action (\ref{actionBS}) within the WDW patch, integrating out the time $t$ coordinate. Here we can ignore the contribution from the boundary as the Gibbons-Hawking term of this boundary turns out to be vanishing. We finally find that the final action is proportional to our complexity functional $I^{tot}_d[\phi,g]$ (\ref{hacd}) with the normalization (\ref{normnnn}) up to surface terms at the AdS boundary $z=0$ due to partial integrations:
\ba
I^{WDW}_d=(d-2)\cdot n_d\cdot I^{tot}_d[\phi,g]+(\mbox{IR Surface Term}), \label{conectwdw}
\ea
where the numerical constant $n_d$ is defined by
\be
n_d=\int^{\pi/2}_{-\pi/2}dt(\cos　t)^{d-2}=\frac{\s{\pi}\Gamma
\left(\f{d-1}{2}\right)}{\Gamma\left(\f{d}{2}\right)}.
\ee
In the above computation, by introducing the Gibbons-Hawking term for
the $d$ dimensional boundary time like surface given by $z=\ep$, the surface terms
on this surface which are produced by the partial integrations of bulk action are all cancelled with the Gibbons-Hawking term. Therefore in the surface terms in (\ref{conectwdw}) is localized at the IR boundary, which is at $z=\infty$ and gives the vanishing contribution for the Poincare AdS coordinate.

For example, when $d=3$ with $h_{ij}=\delta_{ij}$ (setting $x_3=z$), so that it fits with the
Poincare AdS$_4$, we find
\be
\begin{split}
 I^{WDW}_3 = \frac{1}{16\pi G_N}\int^{\infty}_{\ep} dz\int d^2x\int^{\pi/2}_{-\pi/2}dt \bigg[ & 6e^{3\phi}(\cos^3 t-\cos t\cos 2t) \\ & -2e^{\phi}\cos t((\de_i\phi)^2+2\de_i\de
_i\phi)\bigg],
\end{split}
\ee
which reproduces (\ref{conectwdw}) after we integrate $t$ and perform a partial integration with the boundary term at $z=\ep$ cancelled by the Gibbons-Hawking term at $z=\ep$.

When $d=2$ we find
\ba
I^{WDW}_2=\frac{1}{8G_N} \int dz dx[-(\de_1^2+\de_2^2)\phi], \label{wwkk}
\ea
which indeed leads to vanishing action up to partial integrations, where again the boundary term at $z=\ep$ is cancelled by the Gibbons-Hawking term at $z=\ep$. Therefore we simply find
$I^{WDW}_2=-\frac{1}{8G_N}\int dx (\de_z \phi)_{z=\infty}$, where note that $z=\infty$ is the IR boundary.  Since we have $\phi=-\log z$ and $\phi=-\log\sinh z$ for Poincare and global AdS$_3$,
we get
\ba
&& I^{WDW}_2=0\ \ \ \mbox{for CFT$_2$ vacuum on R$^1$ dual to the Poincare AdS$_3$},\no
&& I^{WDW}_2=\frac{\pi}{4G_N}\ \ \ \mbox{for CFT$_2$ vacuum on S$^1$ dual to the Poincare AdS$_3$}.\nonumber
\ea
Interestingly, this agrees with the evaluation of holographic complexity with the prescription in \cite{Reynolds:2016rvl}.

The above relation shows that there is no difference with respect to the equation of motion for $\phi$ between the ``Complexity = Action'' approach and our proposal. However in the $d=2$ limit they differs significantly due to $(d-2)$ factor in (\ref{conectwdw}). In our proposal, the complexity functional for 2D CFTs is obtained as $\lim_{d\to 2}(I^{tot}_{d}[\phi,g]-I^{tot}_{d}[0,g])
=\lim_{d\to 2}\left[(I^{WDW}_{d}-I^{WDW}_{d}|_{\phi=0})/(d-2)\right]$, which coincides with the Liouville action $I_L[\phi,g]$. On the other hand, there are no bulk contributions in
$I^{WDW}_2$ as clear from (\ref{wwkk}). This is essential reason why the former have the UV divergence $O(\ep^{-1})$, while the latter does not.

\subsection{Higher Derivative Terms and Anomalies} \label{hdanom}

In our optimization of two dimensional CFTs, we minimized the overall factor of wave functional, which is the same as the partition function $Z$ for the region $\ep<z<\infty$. The Liouville action which we minimize is given by the log of this partition function $S_L=\log Z$. Therefore even for higher dimensional CFTs one may naively suspect that the complexity functional $I_d$ may also be written as $I_d=\log Z_d$ for $d$-dimensional CFTs. This indeed works for $d=3$ as the UV divergent terms produces the two terms in the action (\ref{hacd}). The situation is different for $d=4$ due to the presence of conformal anomaly \cite{Rieg} and we need to have forth derivative terms in addition to the action (\ref{hacd}). As we have explained in appendix \ref{SZcase}, here we just mention the final form of $I_4$ that correctly reproduces the anomalies in a four dimensional CFT,
\be \label{s4d1cov1}
\begin{split}
I_{4} = \int d^4x\sqrt{g} \bigg[&\alpha_1 + \alpha_2 (\partial^{\mu}\phi \partial_{\mu}\phi) + \alpha_3 (\partial^{\mu}\phi \partial_{\mu}\phi)^2
+ \alpha_4 (\nabla^{\mu}\partial_{\mu} \phi)^2 \\& + \alpha_5(\nabla^{\mu}\partial_{\mu} \phi)(\partial^{\mu}\phi \partial_{\mu}\phi) \bigg],
\end{split}
\ee
where the terms with corresponding coefficients $\alpha_3,~\alpha_4,~\alpha_5$ denote the fourth derivative terms, responsible for producing correct anomalies. It should be mentioned that here we only consider metrics which are of the Weyl scaling type \eqref{met},
\be \label{metcovadsa5}
g_{\mu\nu} = e^{2 \phi}~ h^\text{flat}_{\mu\nu},
\ee
with $h^\text{flat}_{\mu\nu}$ corresponding to Euclidean flat space.

In appendix \ref{SZcase}, we firstly explain in detail how this action in  (\ref{s4d1cov1}) produces the correct anomalies for four dimensional CFT. Next we also explain how the equations of motion following from this action allows time slice of $AdS_5$, i.e. hyperbolic space $H_4$, as a solution.

Here we should admit that the action (\ref{s4d1cov1}) is not bounded from below and hence cannot be minimized, therefore we can only extremize it. In this aspect, the action (\ref{hacd}) without higher derivatives as we assumed in section \ref{2dcftopt} has an advantage over the modified action we are discussing here.

\section{Discussions: Time Evolution of TFD States and Phase Transitions}\label{DiscussionTFD}

So far we have studied stationary quantum states in CFTs. For further understandings of the dynamics of CFTs, we would like to turn to time dependent states in this section, focusing on 2D CFTs for simplicity. In particular we consider a simple but non-trivial class of time-dependent states, given by the time evolution of thermo field double states (TFD states) in 2D CFTs:
\be
|TFD(t)\lb =\frac{1}{\s{Z_{\beta}(t)}}\sum_{n}e^{-\frac{\beta}{4}(H_1+H_2)}e^{-it(H_1+H_2)}|n\lb_1|n\lb_2,
\ee
where the total Hilbert space is doubled $H_{tot}=H_1\otimes H_2$ ($H_1$ is the original CFT Hilbert space and $H_2$ is its identical copy). Its density matrix\footnote{However note that $\rho(t)$ can not be obtained from the analytic continuation $\tau=it$ of Euclidean TFD density matrix
$\rho(\tau)=|TFD(\tau)\lb \la TFD(\tau)|$ defined by the Euclidean path-integral for the Euclidean time region $-\beta/4-\tau\leq \tau\leq \beta/4+\tau$. Instead it is obtained from
$\rho'(\tau)=|TFD(\tau)\lb \la TFD(-\tau)|$.} is given by $\rho(t)=|TFD(t)\lb \la TFD(t)|$ and if we trace out $H_2$, then the reduced density matrix $\rho_1$ is time-independent, given by the standard canonical distribution $\rho_1\propto e^{-\beta H_1}$. However the TFD state $|TFD(t)\lb$ shows very nontrivial time evolution and is closely related to quantum quenches as pointed out in \cite{HaMa}.

\subsection{Optimizing TFD State}

\begin{figure}[b!]
  \centering
  \includegraphics[width=7cm]{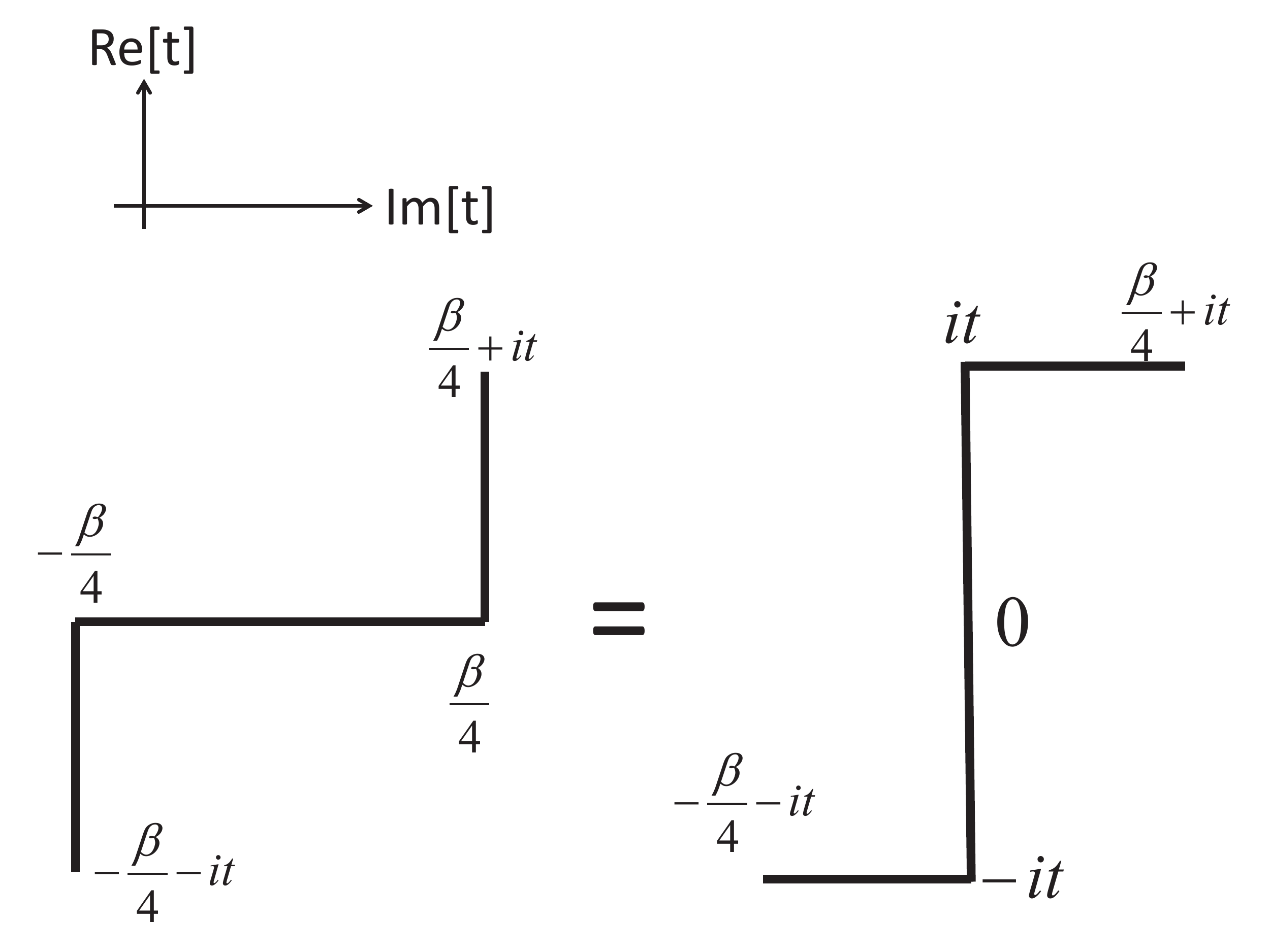}
  \caption{The choices of path-integral contour for the TFD states. We employ the time coordinate
  in the first CFT $H_1$. The left and right choices are equivalent.}
\label{TFDpath}
  \end{figure}

Motivated by this, let us study the path-integral expression of $|TFD(t)\lb$. First we can create the initial TFD state $|TFD(0)\lb$ by the Euclidean path-integral for the range of Euclidean time $\tau$:
\be
-\f{\beta}{4}\leq \tau \leq \frac{\beta}{4}.
\ee
After this path-integration, we can perform the Lorentzian path-integral by $it$ on each CFT.
This integration contour is depicted as the left picture in Fig.\ref{TFDpath}.
However, as we will see later, there is an equivalent but more useful contour given by the right picture in Fig.\ref{TFDpath}. This is because we can exchange the Euclidean time evolution $e^{-\beta(H_1+H_2)/4}$ with the real time one $e^{-it(H_1+H_2)}$.

Now we consider an optimization of this path-integral. For the Euclidean part we can apply the same argument as before and minimize the Liouville action. Next we need to consider an optimization of the real time evolution. However, we would like to argue that this Lorentzian path-integral cannot be optimized. A heuristic reason for this is that if the final state even after a long time evolution,
is still sensitive to the initial state as opposed to the Euclidean path-integral. On the other hand, if we perform an Euclidean time evolution for a period $\Delta \tau$, then the final state is insensitive to the high momentum mode $k\gg 1/\Delta \tau$ of the initial state. Once we assume this argument, we can understand the reason why we place the Lorentzian time evolution in the middle sandwiched by the Euclidean evolution as this obviously reduces the value of $S_L$.
It is an intriguing future problem to verify these intuitive arguments using the tensor network framework.

\begin{figure}[h!]
  \centering
  \includegraphics[width=7cm]{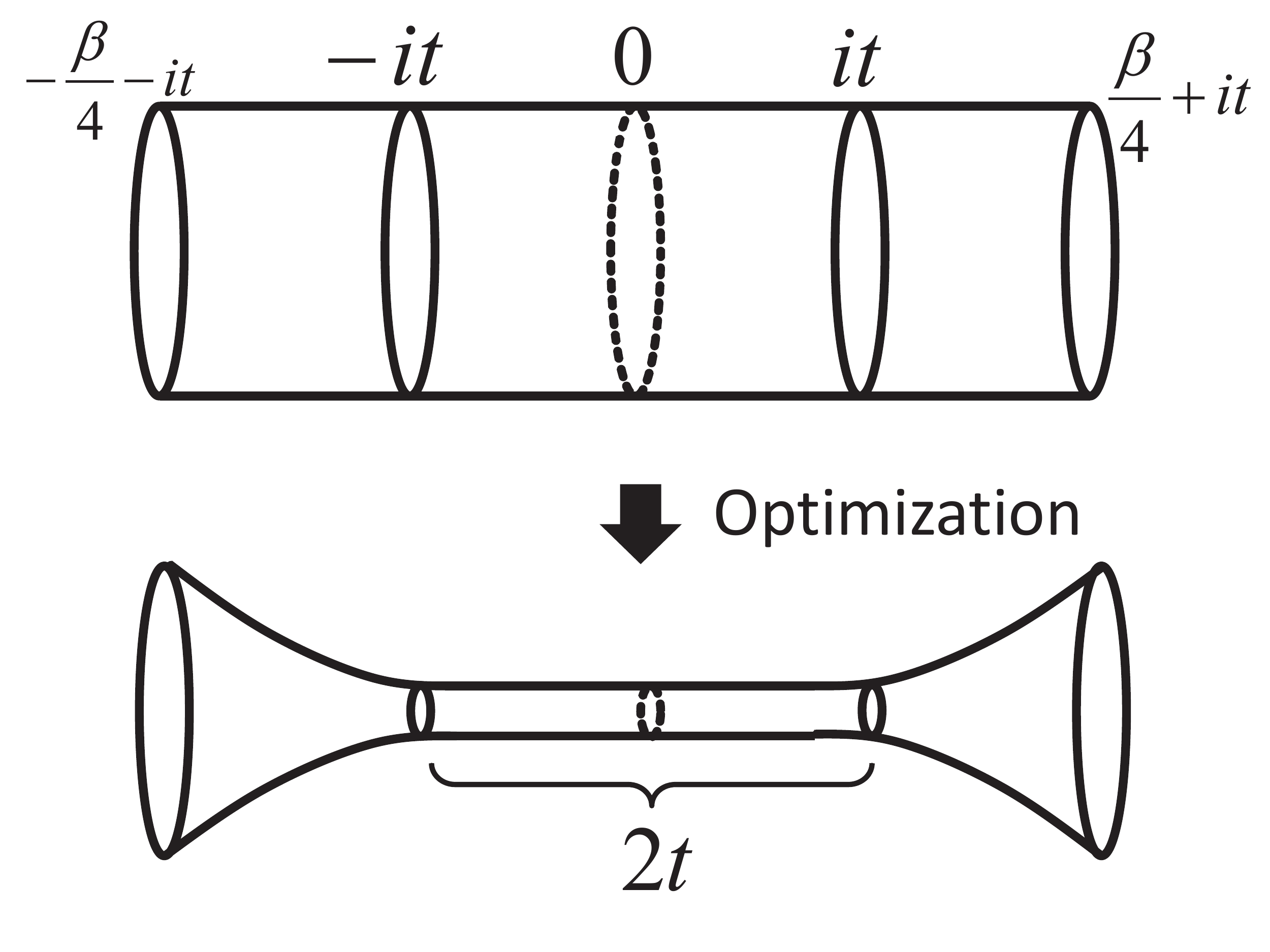}
  \caption{The optimization of path-integral for the TFD states. The Euclidean path-integral is optimized by minimizing the Liouville action. We assumed that the Lorentzian one cannot be optimized.}
\label{TFDopt}
  \end{figure}

Assuming that the above prescription of optimization is correct at least semi-quantitatively, we can find the following solution (remember we set $z=-\tau$):
 \ba
 e^{2\phi}=
 \Biggl\{
 \begin{array}{l}
   \frac{4\pi^2}{\beta^2}\cos^{-2}\left(\f{2\pi \Re[z]}{\beta}\right),\ \ \ \ \ \
   (-\f{\beta}{4}-it<z<-it,\ \ \ \  it<z<it+\frac{\beta}{4})    \\
   \frac{4\pi^2}{\beta^2},\ \ \ \ \ \
   (-it<z<it).  \label{timesol}
 \end{array}
 \ea
This is depicted in Fig.\ref{TFDopt}.
  
It is also intriguing to estimate the complexity. For the Euclidean part, we proposed that it is given by the Liouville action as we explained before. For Lorentzian part, there is no obvious candidate. However since we assumed that $\phi$ is constant during the real time evolution, we can make a natural identification: the Liouville potential term gives the complexity. This is clear from the fact that the complexity should be proportional to the number of tensors.\\
Thinking this way, we find
\be
S_L(t)=S_L(0)+\left(\f{2\pi}{\beta}\right)^2\cdot\frac{c}{6}\cdot t.
\ee
This linear $t$ growth is consistent with the basic idea in \cite{Susskind}.
Since the energy in our 2D CFT at  finite temperature $T=1/\beta$ is given by
\be
E_{CFT}=\frac{\pi^2}{3}cT^2,
\ee
we find
\be
\f{dS_L(t)}{dt}=2E_{CFT}. \label{gtwqlaj}
\ee
Interestingly this growth is equal to a half of the gravity action $I_{WDW}$ on the WDW patch for holographic complexity found in \cite{BrownSusskind1,BrownSusskind2}, where the holographic complexity $C_{A}$ is conjectured to be
$C_{A}=\frac{I_{WDW}}{\pi}$ (\ref{CAcon}). Note that we are shifting both the time in the first and second CFT at the time time.
This relation $\f{dI_{WDW}}{dt}=2\f{d S_L}{dt}$ may be natural because the partition function of CFTs $Z\sim e^{\cal A}$ is the square of that of wave functional in CFTs $|\Psi|^2\sim e^{2S_L}$.

It is also intriguing to consider a pure state which looks thermal when we coarse-grain its total system. One typical such example in CFTs is obtained by regularizing a boundary state $|B\lb$
\be
|\Psi_B\lb={\cal N}_B e^{-\beta H/4}|B\lb,
\ee
where ${\cal N}_B$ is the normalization such that $\la \psi_B|\psi_B\lb=1$. This can also be regarded as an approximation of global quenches \cite{GQ,TaUg}.
This quantum state is dual to a single-sided black hole \cite{HaMa}
shows the linear growth of holographic entanglement entropy which matches with the 2D CFT result in
\cite{GQ}.  This state after our path-integral optimization is clearly given by a half of TFD (\ref{btzwq}) for $0<z<\beta/4-\ep$. The boundary at $z=0$ corresponds to that of the boundary state
$|B\lb$ which matches with the AdS/BCFT formulation \cite{Ta}. Thus the growth of the complexity functional is simply given by a half of the TFD case (\ref{gtwqlaj}).

\subsection{Comparison with Eternal BTZ black hole}

The time evolution of TFD state provides an important class of time-dependent states and
here we would like to discuss possible connections between our optimization procedure and its gravity dual given by the eternal BTZ black hole. In this section we set $\beta=2\pi$ for simplicity.

First let us try to assume that the dual geometry for this time-dependent quantum states has a property that each time slice is given by a space-like geometry which is a solution to Liouville equation. Any solution to the Liouville equation is always a hyperbolic space with a constant curvature. Such a hyperbolic space at each time $t$ is obtained by taking a union of all geodesic which connects two points at the time $t$ with the same space coordinate in the two different boundaries, given explicitly by
\ba
&& ds^2=e^{2\phi(z)}(dz^2+dx^2),  \no
&& e^{2\phi(z)}=\frac{\left(\frac{1}{\cosh t}\right)^2}{\sin^2\left(\frac{z}{\cosh t}\right)}.  \label{werq}
\ea
By the transformation
\be
\cosh\rho=\frac{\frac{1}{\cosh t}}{\sin\left(\frac{z}{\cosh t}\right)},
\ee
the metric is rewritten as
\be
ds^2=\cosh^2\rho\left(dx^2+\frac{\sinh^2\rho}
{\sinh^2\rho\cosh^2\rho+\tanh^2 t\cosh^2\rho}d\rho^2\right).
\ee
Indeed the whole BTZ metric
\be
ds^2=-\sinh^2\rho dt^2+d\rho^2+\cosh^2\rho dx^2,
\ee
can be rewritten into the metric
\be
ds^2=\frac{1}{\sin\left(\frac{z}{\cosh\tau}\right)^2\cdot \cosh^2\tau}
\left[-d\tau^2+dx^2+\left(z\tanh\tau (d\tau)-dz\right)^2\right],
\ee
via the coordinate transformation
\ba
\cosh\rho=\frac{1}{\cosh\tau\cdot \sin\left(\frac{z}{\cosh\tau}\right)},
\ \ \ \ \tanh t=\frac{\tanh\tau}{\cos\left(\frac{z}{\cosh\tau}\right)}.
\ea

However if we evaluate its action (as in the computation of (\ref{btzfind}))
we find (we recovered $\beta$ dependence)
\ba
S_L=\frac{8\pi}{\ep}-\frac{4\pi^3}{\beta\cosh t}.
\ea
Thus there is no linear $t$ growth. In this way, this surface does not seem to have an expected
property which supports the linearly growing complexity argued in many papers \cite{SUR,Susskind,BrownSusskind1,BrownSusskind2,Lehner:2016vdi}.

Now we would like to turn to another candidate: maximal time slice, whose volume was conjectured to be one candidates of holographic complexity \cite{SUR,Susskind}. Note that this maximal time slice is not a solution to Liouville equation as opposed to the previous hyperbolic space (\ref{werq}), which is constructed from geodesics.

The BTZ metric behind the horizon can be obtained by the analytic continuation
$\rho=i\kappa$, $\ti{t}=t+\f{\pi i}{2}$.
Maximal volume surface with boundary time $t$ is determined by the equation
\be
s^2(t)=\f{\cosh^2\rho \sinh^4\rho}{\dot{\rho}^2-\sinh^2\rho}=\f{\cos^2\kappa \sin^4\kappa}{-\dot{\kappa}^2+\sin^2\kappa}.
\ee
$s(t)^2$ increases monotonically as $t~(\geq 0)$ increases, with boundary value $s(0)=0$ and $s(\infty)=1/2$.
The induced metric on the maximal volume time-slice is
\be
ds^2=\cosh^2\rho \left[\f{\sinh^2\rho}{s(t)^2+\sinh^2\rho \cosh^2\rho}d\rho^2+dx^2\right].
\ee
The curvature of the maximal volume time slice is not constant, therefore the time slice is not
 hyperbolic. Then, we find that the volume term increases linearly in time. Finally we obtain
  \be
 \f{c}{24\pi}\f{d({\rm Vol(t)})}{dt}\approx \f{c}{12},
  \ee
at late time (here we used the same normalization as our proposal for the Liouville action). This behavior is in contrast to the previous hyperbolic time slice, where the action approaches monotonically to some constant value.

In summary, the above arguments imply that for a generic time dependent background, the assumption that a preferred time slice in a gravity dual is described by Liouville equation, is not compatible with the requirement that the Liouville action gives a measure of complexity. Thus an extension of our proposal in this paper to time-dependent backgrounds looks highly non-trivial and deserves future careful studies.

\subsection{Comment on Phase Transition}

\begin{figure}[b!]
  \centering
  \includegraphics[width=8cm]{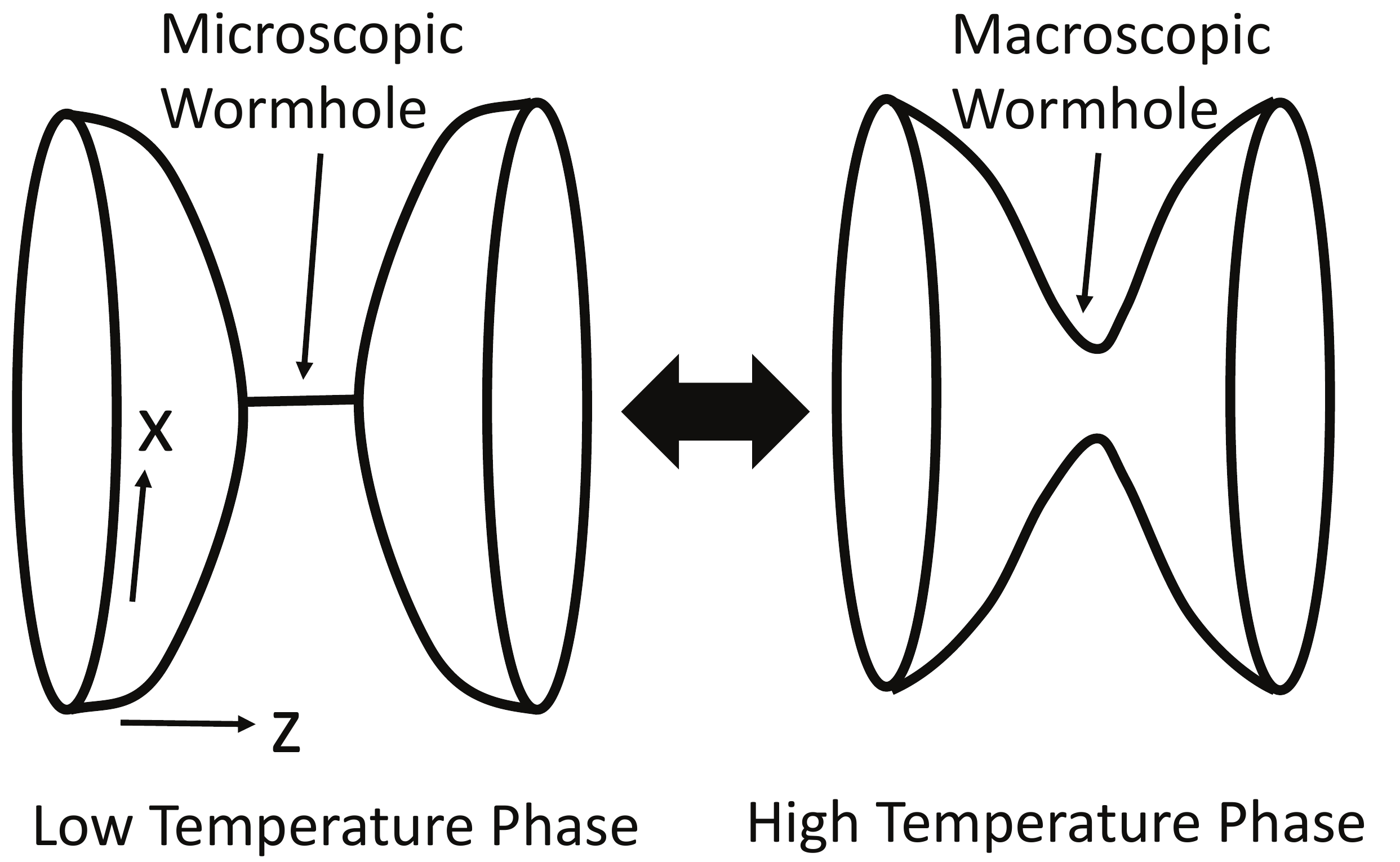}
  \caption{The expected description of confinement/deconfinement phase transition in our optimization of path-integral for the TFD states. At low temperature the two CFTs are connected through a microscopic bridge with entanglement entropy $O(1)$ in the tensor network. At high temperature the bridge gets macroscopic and has entanglement entropy $O(c)$.}
\label{fig:phase}
  \end{figure}

It is also intriguing to discuss how we can understand the confinement/deconfinement phase transition in our approach. For this, we focus on the initial state $|TFD(t=0)\lb$. Since our approach is based on pure states we need to consider the wave functional of TFD state (at temperature $T$) and see how the corresponding tensor network changes as a function of $T$. It is obvious that at high temperature, the connected network which looks like macroscopic wormhole is realized and this should be described by the optimized path-integral on the Einstein-Rosen bridge (\ref{phforbtz}). As we make the temperature lower, the neck of bridge gets squeezed and eventually disconnected in a macroscopic sense. Here we mean by the macroscopic the quantum entanglement of order $O(c)=O(1/G_N)$. Refer to Fig.\ref{fig:phase}.

Since the TFD state has non-zero (but sub-leading order $O(1)$) entanglement entropy between the two identical CFTs even at low temperature, there should be a microscopic bridge or wormhole (following ER=EPR conjecture \cite{EREPR}) which connects the two sides in the tensor network description. In this low temperature, the bridge is due to the singlet sector of the gauge theory and is in its confined
phase. In large $c$ holographic CFTs, there should be a phase transition of the macroscopic form of the tensor network at the value $\beta=1/T=2\pi$ predicted by AdS$_3/$CFT$_2$.
Naturally, we expect that the favored phase of a given quantum state $\Psi$ is the one which has smaller complexity $C_\Psi$.

However, in the current form of our arguments based on the path-integral optimization, it is not straightforward to compare the value of the complexity (i.e. Liouville action) for the confinement/deconfinement phase transition. This is because we can only define the difference of
complexity which depends on the reference metric. In this phase transition, the topology of the reference space changes and it is difficult to know how to compere them precisely.

Nevertheless, it might be useful to try to roughly estimate the complexity. For this we assume that the complexity for the deconfined phase (denoted by $C_{dec}$) is estimated by by the bridge solution (\ref{btpp}) and that for the confined phase (denoted by $C_{con}$) is by the twice of the vacuum result (\ref{optijwidww}), which leads to
\be
C_{dec}\approx \frac{c}{3\ep}-\frac{\pi^2 c}{6\beta},\ \ \  \ C_{con}\approx \frac{c}{3\ep}-\frac{c}{3}.
\ee
Qualitatively, this has an expected behavior that $C_{dec}<C_{con}$ for $\beta \ll 2\pi$ and vise versa, though the phase transition temperature reads $\beta=\pi^2/2$, which is slightly different from the gravity result $2\pi$.

Another interesting interpretation of the phase transition can be found from a property in the Liouville CFT. It is known that the (chiral) conformal dimension $h$ of any local operators in Liouville theory  has an upper bound (so called Seiberg bound \cite{Se}):
\be
h \leq \frac{c-1}{24}, \label{sebgfd}
\ee
which implies the non-normalizability of the corresponding state.
The operator which violates the bound should be regarded as a (normalizable) quantum state.
In the large $c$ limit, this bound (\ref{sebgfd}) agrees with the condition that the conical deficit angle parameter $a$ given by (\ref{three}), takes a real value, for which the metric is that for confined phase. When it is violated, $a$ becomes imaginary and the metric changes into that for the deconfined phase (Einstein-Rosen bridge). This behavior seems to fit very nicely with the gravity dual prediction and to proceed this further is an important future problem.

\section{Conclusions}\label{Concl}

In this work, we proposed an optimization procedure of Euclidean path-integrals for quantum states in CFTs. The optimization is described by a change of the background metric on the space where the path integral is performed. The optimization is completed by minimizing the complexity functional $I_\Psi$ for a given state $|\Psi\lb$, which is argued to be given by the Liouville action for 2D CFTs. The Liouville field $\phi$ corresponds to the Weyl scaling of the background metric. Since this complexity is defined from Euclidean path-integrals, we propose to call this ``Path-Integral Complexity''.

Through calculations in various examples in 2d CFTs, we observed that optimized metrics for static quantum states coincide with those of time slices of their gravity duals. Thus we argued that our path-integral optimization offers  a continuous version of the tensor network interpretation of AdS$_3$/CFT$_2$ correspondence. Moreover, we also find a simple formula to calculate the energy density for each quantum state.

At the same time, we provide a field theory framework for evaluating the computational complexity of any quantum states in CFTs. Note however, that in 2D CFTs, due to the conformal anomaly, the complexity functional (i.e. Liouville action) depends on the reference metric. Therefore, we proposed to use the difference of the action, which is expected to give a relative difference of complexity between the optimized network and the initial un-optimized one. We evaluated this quantity in several examples.

In order to calculate the entanglement entropy, we studied an optimization of reduced density matrices. After the optimization we find that the geometry is given by two copies of entanglement wedge and this nicely fits into the gravity dual. The entanglement entropy is finally reduced to the length of the boundary of the entanglement wedge and precisely reproduces the holographic entanglement entropy.

Even though in most parts of this paper our analysis is devoted to static quantum states, we also discussed how our optimization of path-integrals can be applied to time-dependent backgrounds in 2D CFTs. Especially, we considered the time evolution of thermo-field states which describe finite temperature states as a basic example. Our heuristic arguments show that an wormhole throat region linearly grows under the time evolution, which is consistent with holographic predictions. Moreover, we discussed how to interpret the confinement/deconfinement phase transition in terms of tensor networks and our path-integral approach, whose details will be an interesting future problem. However, a precise connection between Liouville action and time-dependent states in 2D CFTs is still not clear and this was left as an important future problem.

In the latter half of this paper, we investigated the application of our optimization method to CFTs in other dimensions than two. In one dimension, we find that 1D version of Liouville action naturally arises from the conformal symmetry breaking effect in NAdS$_2/$CFT$_1$ and this explains the emergence of extra dimension as in the AdS$_3/$CFT$_2$ case. 

In higher dimensions, we expect that the optimization procedure gets very complicated as we need to change not only the scaling mode but also other components of the metric as opposed to the 2D case. We focused on the Weyl scaling mode and proposed a complexity functional which looks like a higher dimensional version of Liouville action. However, notice the crucial difference from the 2D case that the higher dimensional action does not depend on the reference metric. We confirmed that this reproduces the correct time slice metric for the vacuum states and correct holographic entanglement entropy when the subsystem is a round sphere. We pointed out an interesting direct connection to earlier proposal of holographic complexity \cite{BrownSusskind1, BrownSusskind2}, which may suggest we should optimize with respect to all components of the metric. We also analyzed the spherically symmetric excited states and found that the optimized metric agrees with the AdS Schwarzschild one up to the first order contribution of the mass parameter. We observed that for CFTs in any dimensions (including 2D), in order to take into account higher order back-reactions, we need to treat the Liouville mode $\phi$ in a quantum way. It is also possible to include higher derivative corrections without losing the above properties as we discussed in appendix (\ref{SZcase}). One advantage of this is that we can realize the higher dimensional conformal anomaly. However there is also a disadvantage that the action is no longer positive definite and cannot be minimized but extremized. These issues on higher dimensional CFTs should deserve further studies.

Last but not least, our approach based on the optimization of path-integrals is a modest but important step towards understanding of the basic mechanism of the AdS/CFT correspondence. For the future, apart from the questions we already mentioned above, there are many new directions for investigations like e.g. computation of correlation functions, generalizations to non-conformal field theories and understanding a precise connection to AdS/CFT including $1/c$ expansions etc.

\medskip
{\bf \large Acknowledgments}

We thank Bartek Czech, Glen Evenbly, Rajesh Gopakumar, Kanato Goto, Yasuaki Hikida, Veronika Hubeny, Satoshi Iso, Esperanza Lopez, Alex Maloney, Rob Myers, Yu Nakayama, Tatsuma Nishioka, Yasunori Nomura, Tokiro Numasawa, Hirosi Ooguri, Alvaro-Veliz Osorio, Fernando Pastawski, Mukund Rangamani, Shinsei Ryu, Brian Swingle, Joerg Teschner, Erik Tonni, Tomonori Ugajin, Herman Verlinde, Guifre Vidal, Spenta Wadia, Alexander Westphal, Kazuya Yonekura for useful discussions and especially Rob Myers, Beni Yoshida and Alvaro Veliz-Osorio for comments on the draft.  MM and KW are supported by JSPS fellowships. PC and TT are supported by the Simons Foundation through the ``It from Qubit'' collaboration. NK and TT are supported by JSPS Grant-in-Aid for Scientific Research (A) No.16H02182. TT is also supported by World Premier International Research Center Initiative (WPI Initiative) from the Japan Ministry of Education, Culture, Sports, Science and Technology (MEXT). PC, MM, TT and KW thank very much the long term workshop ``Quantum Information in String Theory and Many-body Systems'' held at YITP, Kyoto where this work was initiated. NK would like to acknowledge the hospitality of the theory group at TIFR, Mumbai during an academic visit and for useful discussions during a seminar where parts of this work was presented. TT is very much grateful to the conference ``Recent Developments in Fields, Strings, and Gravity'' in Quantum Mathematics and Physics (QMAP) at UC Davis, the international symposium ``Frontiers in Mathematical Physics'' in Rikkyo U., the conference ``String Theory: Past and Present (Spenta Fest)'' in ICTS, Bangalore, the workshop ``Tensor Networks for Quantum Field Theories II'' at Perimeter Institute, the workshop ``Entangle This: Tensor Networks and Gravity'' at IFT, Madrid, the ``Universitat Hamburg - Kyoto University Symposium''at DESY in Hamburg University, where the contents of this paper were presented.

\appendix

\section{Comparison with Earlier Liouville/3D gravity Relation} \label{ap:3dgravity}

Here we would like to compare our Liouville theory obtained from an optimization of Euclidean path-integrals with the earlier relation \cite{CHD} between 3D gravity and Liouville theory.
For simplicity we set the AdS radius to unit $R=1$ below.
We employ the Chern-Simons description of 3D gravity \cite{WittenCS}, the two $SL(2,R)$ gauge fields $A$ and $\bar{A}$ correspond to the triad $e$ and spin connection $\omega$ via $A=\omega+e$
and $\bar{A}=\omega-e$. If we choose the solution:
\be
A=\left(
  \begin{array}{cc}
    \frac{dr}{2r} & -T_{++}(x^+)\frac{dx^+}{r} \\
    rdx^+ & -\frac{dr}{2r} \\
  \end{array}
\right),\ \ \ \
\bar{A}=\left(
  \begin{array}{cc}
    -\frac{dr}{2r} & r dx^- \\
    -T_{--}(x^-)\frac{dx^-}{r} & \frac{dr}{2r} \\
  \end{array}
\right),
\ee
we obtain a series of solutions which describe gravitational waves on a pure AdS space
(called Banados geometry) \cite{Ban}:
\be
ds^2=\frac{dr^2}{r^2}-\left(r^2+T_{++}T_{--}r^{-2}\right)dx^+dx^-
-T_{++}(dx^+)^2-T_{--}(dx^-)^2. \label{bang}
\ee
This satisfies the equation of motion iff $T_{++}$ and $T_{--}$ are functions of $x^+$ and $x^-$ respectively, i.e. $\de_- T_{++}=\de_{+} T_{--}=0$. If we set $T_{++}$ and $T_{--}$ to be constants,
the geometry becomes a BTZ black hole.

\subsection{Review of Earlier Argument}
In the paper \cite{CHD}, motivated by the asymptotic behavior of BTZ black hole solutions,
the following gauge choices are imposed: $A=(G_1)^{-1}dG_1$ and $A=(G_2)^{-1}dG_2$ (note that there is no bulk degrees of freedom in Chern-Simons gauge theories), where $G_1$ and
$G_2$ are expresses as follows
\ba
G_1=g_1(x^+,x^-)\cdot
\left(
  \begin{array}{cc}
    \s{r} & 0 \\
    0 & \frac{1}{\s{r}} \\
  \end{array}
\right),\ \ \
G_2=g_2(x^+,x^-)\cdot
\left(
  \begin{array}{cc}
    \frac{1}{\s{r}} & 0 \\
    0 & \s{r} \\
  \end{array}
\right).
\ea
In the above expression $g_1$ and $g_2$ are $SL(2,R)$ matrices and describe the boundary degrees of freedom. Note that we can show
\ba
A_r=-\bar{A}_r\left(
  \begin{array}{cc}
    \frac{1}{2r} & 0 \\
    0 & -\frac{1}{2r} \\
  \end{array}
\right), \
A_{\pm}=\left(
  \begin{array}{cc}
    a^{(3)} & a^{(+)}/r \\
    ra^{(-)} & -a^{(3)} \\
  \end{array}
\right),
\bar{A}_{\pm}=\left(
  \begin{array}{cc}
    \bar{a}^{(3)} & \bar{a}^{(+)}/r \\
    r\bar{a}^{(-)} & -\bar{a}^{(3)} \\
  \end{array}
\right), \nonumber
\ea
where we defined $a_{\pm}=(g_1)^{-1}\de_+ g_1$ and $\bar{a}_{\pm}=(g_2)^{-1}\de_- g_2$.
Next we impose the chiral gauge choices $a_{-}=\bar{a}_{+}=0$. In this case the gauge theory for
$A$ and $\bar{A}$ becomes equivalent to the chiral and anti-chiral $SL(2,R)$ WZW model, respectively \cite{CHD}. Thus, by combining $g_1$ and $g_2$ as $g=g_1^{-1}g_2$ we obtain a $SL(2,R)$ WZW model.
If we describe the $SL(2,R)$ group element by
\ba
g=\left(
  \begin{array}{cc}
   1  &  X   \\
   0  &  1   \\
  \end{array}
\right)
\left(
  \begin{array}{cc}
   e^{\phi}  &  0   \\
   0  &  e^{-\phi}   \\
  \end{array}
\right)
\left(
  \begin{array}{cc}
  1   &  0   \\
  Y   &  1   \\
  \end{array}
\right),
\ea
then the WZW model is described by the action
\be
S_{WZW}=\int dx^+dx^{-}\left[2\de_+\phi\de_{-}\phi+2e^{-2\phi}(\de_- X)(\de_+ Y) \right].
\ee
We can find the solutions to the equation of motion for $S_{WZW}$ such that
\ba
&& \de_+ Y=-\ap(x^+)\cdot e^{2\phi}, \ \ \de_- X=\beta(x^-)\cdot e^{2\phi}, \no
&& \de_+\de_{-}\phi=\ap(x^+)\beta(x^-)e^{2\phi}. \label{qqw}
\ea
Now we set $\ap(x^+)=\beta(x^-)=1$ via a coordinate transformation. Note that the final equation in
(\ref{qqw}) coincides with the equation of motion of Liouville theory and this provides the connection between the 3D gravity and Liouville theory.  Finally, the gauge field $A$ and $\bar{A}$ for this solution read
\ba
A=\left(
  \begin{array}{cc}
    \frac{1}{2r} & \left(-\de^2_+\phi+(\de_+\phi)^2\right)\frac{dx^+}{r} \\
    rdx^+ & -\frac{1}{2r} \\
  \end{array}
\right), \ \ \
\bar{A}=\left(
  \begin{array}{cc}
    -\frac{1}{2r} & rdx^- \\
    \left(-\de^2_-\phi+(\de_-\phi)^2\right)\frac{dx^-}{r} & \frac{1}{2r} \\
  \end{array}
\right). \nonumber
\ea
Thus we find that the serious of the above solutions correspond to the Banados geometry
(\ref{bang}) with the
energy stress tensor in the Liouville CFT:
$T_{\pm \pm}=\de^2_{\pm}\phi-(\de_{\pm}\phi)^2$.

\subsection{Comparison with Our Approach}

Now let us compare the above earlier argument to our metric ansatz (\ref{ourm}), which fits naturally with our path-integral optimization argument. We work in Euclidean signature and consider the Euclidean version of Banados metric (\ref{bang}) given by
\be
ds^2=\frac{dz^2}{z^2}+\left(z^2+T(w)\bar{T}(\bar{w})z^{-2}\right)dwd\bar{w}
+T(w)dw^2+T(\bar{w})d\bar{w}^2. \label{bangl}
\ee
This metric is mapped into the standard Poincare  AdS$_3$ metric
$ds^2=\frac{d\eta^2+dx^2+d\tau^2}{\eta^2}$ via
\ba
&& \tau+ix=A(w)-\frac{z^2A'(w)B''(\bar{w})}{4A'(w)B'(\bar{w})+z^2A''(w)B''(\bar{w})},\no
&& \tau-ix=B(\bar{w})-\frac{z^2B'(\bar{w})A''(w)}{4A'(w)B'(\bar{w})+z^2A''(w)B''(\bar{w})},\no
&& \eta=\frac{4z(A'(w)B'(\bar{w}))^{3/2}}{4A'(w)B'(\bar{w})+z^2A''(w)B''(\bar{w})}.
\label{coraa}
\ea
Here $A$ and $B$ are holomorphic and anti holomorphic functions, respectively and the
energy stress tensors are expresses as
\be
T(w)=\frac{3A''(w)^2-2A'(w)A'''(w)}{4A'(w)^2}, \ \ \
\bar{T}(\bar{w})=\frac{3B''(\bar{w})^2-2B'(\bar{w})B'''(\bar{w})}{4B'(\bar{w})^2}.
\label{emtw}
\ee

On the other hand, the metric (\ref{ourm}) with the general solution to the Liouville equation
\be
e^{2\phi}=\frac{4A'(y)B'(\bar{y})}{(A(y)+B(\bar{y}))^2}, \label{lihgse}
\ee
is mapped into the same Poincare AdS$_3$ via the map:
\be
\sinh\rho=\frac{\tau}{\eta},\ \ \ A(y)+B(\bar{y})=2\s{\tau^2+\eta^2},\ \ \
A(y)-B(\bar{y})=2ix.  \label{corbb}
\ee
Note that the energy stress tensor for the Liouville field (\ref{lihgse}) agrees with (\ref{emtw}) as it should be.

Therefore, by combining (\ref{coraa}) and (\ref{corbb}) we obtain a coordinate transformation between the Banados metric $(z,w,\bar{w})$ and our metric $(\rho,y,\bar{y})$. Notice that the map is
 trivial near the AdS boundary such that
$y=w+O(z^2)$ and $\bar{y}=\bar{w}+O(z^2)$ when $z$ is very small.

\section{Holographic Complexity in the Literature} \label{holcomlit}
As mentioned in section \ref{evholcomp}, in this appendix we will consider both CV and CA-conjectures for the computation of holographic complexity and will explicitly determine them for some specific cases like Poincare and global AdS in order to compare them with our set-up.

In what follows we will summarize the behavior of holographic complexity in different situations and with both the CV and CA conjectures \footnote{In this appendix, for the sake of convenience, we are using a convention where we put the AdS radius $R_{AdS}=1$.}.
\begin{enumerate}
\item{Poincare $AdS_{d+1}$} : From \cite{Reynolds:2016rvl}, where the complexity action $I_{WDW}$ is evaluated with the null boundary term found in \cite{Lehner:2016vdi}, we see
\be
\begin{split}
\text{CV conjecture :}&~~ C_V = {V_x \over (d-1)G_N} {1 \over \epsilon^{d-1}}\\
\text{CA conjecture :}&~~ C_A = {I_{WDW} \over \pi}, \\
&~~ I_{WDW} = {4 V_x \over 16 \pi G_N} ~ \log (d-1) {1 \over \epsilon^{d-1}}
\end{split}
\ee
with $V_x=$Volume of the $(d-1)$-dim spatial extent of $CFT_{(d-1)}$.
\item{Global $AdS_{d+1}$} : Following the construction in \cite{CMR} (see appendix C for details) for global $AdS_{d+1}$, we note that the leading divergence in $C_A$ behaves as
\be
C_A \sim {\Omega_{d-1} \over  16 \pi^2 G_N} \log \left( {1 \over \sqrt{\alpha}~ \epsilon}\right) {1 \over \epsilon^{d-1}} + \cdots
\ee
where $\epsilon$ is the UV cut-off  and $\Omega_{d-1}$ being the volume of unit sphere $\mathcal{S}^{d-1}$. The sub-leading contributions include terms starting from $1/\epsilon^{d-1}$, but strikingly enough the leading term has an additional and stronger logarithmic divergence. As explained in \cite{CMR} this comes from one of the joint contributions but suffers from the ambiguity of a parametrization of the null boundary of the WDW patch, and is denoted by the free parameter $\alpha$. In \cite{Reynolds:2016rvl}, a prescription to resolve this ambiguity was proposed and following their construction we see
\be
\begin{split}
\text{CV conjecture :}&~~ C_V = {\Omega_{d-1} \over  G_N} \int_{0}^{\theta_{cut}} {d\theta \over \cos \theta} \tan^{d-1} \theta\\
\text{CA conjecture :}&~~ C_A =  {I_{WDW} \over \pi  }, \\
&~~ I_{WDW} = {1 \over 16 \pi G_N}\bigg[-4 \Omega_{d-1} \int_{0}^{\theta_{cut}} {dt'} \tan^{d} t' \\
&~~~~~~ + 4 \Omega_{d-1} \left( \ln (d-1) + {1 \over d-1}\right) \tan^{d-1} \theta_{cut} \bigg]
\end{split}
\ee
with $\theta_{cut} = \pi/2 - \epsilon$. For some explicit cases, we see that
\be \label{cvcalitgads}
\begin{split}
\text{d=2}&~ \Rightarrow ~  C_V ={\Omega_{1} \over  G_N}\left(\frac{1}{\epsilon }-1 \right),~~~~I_{WDW} ={\Omega_{1} \over 8G_N}\\
\text{d=3}&~ \Rightarrow ~~  C_V ={\Omega_{2} \over  G_N}\left(\frac{1}{2 \epsilon ^2}-\frac{1}{2}\log (2/\epsilon)-\frac{1}{12} \right) , \\ & ~~~~~~~~I_{WDW} ={\Omega_{2}\over 16 \pi G_N}\left[\frac{4\log 2}{\epsilon ^2}-4 \log (\epsilon )- \frac{8\log 2}{3} \right] , \\
\text{d=4}&~ \Rightarrow ~  C_V ={\Omega_{3} \over  G_N}\left(\frac{1}{3 \epsilon ^3}-\frac{5}{6 \epsilon }+\frac{2}{3}\right) ,\\ & ~~~~~~~~I_{WDW} = {\Omega_{3}\over 16 \pi G_N}\left(\frac{4 \log 3}{\epsilon ^3}+\frac{4-4 \log 3}{\epsilon }-2 \pi \right).
\end{split}
\ee
\end{enumerate}

\section{Higher Derivatives in Complexity Functional and Anomalies}\label{SZcase}

As was mentioned in section \ref{hdanom}, in this appendix we would like to explore the possibility of working with complexity functional $I_d$ such that it correctly produces the anomalies for even dimensional CFTs and hence can be considered as the partition function $I_d=\log Z_d$ for $d$-dimensional CFTs.

 Motivated by this, we analyze the AdS$_5/$CFT$_4$ case assuming the relation $I_4=\log Z_4$. We will confirm that the equation of motion for the new action again produces the hyperbolic time slice $H_4$ and moreover its first order perturbation agrees with the AdS$_5$ Schwarzschild black hole solution. The possibility of having extra higher derivative terms in the action functional can be related to the trace anomaly in $CFT_4$
\be  \label{tranom}
\delta_\phi I_{4} = \int d^4x \sqrt{g} ~\phi \left( cW^2 -a E_4 + b \nabla^{\mu} \nabla_{\mu} R \right)
\ee
where $W^2$ is the square of Weyl tensor and $E_4 (= R_{\mu\nu\rho\sigma}^2 -4 R_{\mu\nu}^2 +R^2)$ is the topological Euler density in $4$-dimensions and $\mu=z,x_1,x_2,x_3$. Also, note the last term can be taken care of through a local counter term, see \cite{Rieg}.

As mentioned before, we restricts ourselves here only to the metrics which are of the Weyl scaling type \eqref{met},
\be \label{metcovads5}
g_{\mu\nu} = e^{2 \phi}~ h^\text{flat}_{\mu\nu}.
\ee
with $h^\text{flat}_{\mu\nu}$ corresponding to Euclidean flat space.  It can be shown that the action $I_{4}$, which correctly reproduces \eqref{tranom},  becomes
\be \label{s4d1cov}
\begin{split}
I_{4} =\int d^4x\sqrt{g} \bigg[& \alpha_1 + \alpha_2 (\partial^{\mu}\phi \partial_{\mu}\phi) + \alpha_3 (\partial^{\mu}\phi \partial_{\mu}\phi)^2 \\ &
+ \alpha_4 (\nabla^{\mu}\partial_{\mu} \phi)^2+ \alpha_5(\nabla^{\mu}\partial_{\mu} \phi)(\partial^{\mu}\phi \partial_{\mu}\phi) \bigg]
\end{split}
\ee
such that $\alpha_3 = 6a-3b, ~\alpha_4 =-3b , ~\alpha_5 =-4a+6b$ and $g$ is the determinant of the metric $g_{\mu\nu}$ in \eqref{metcovads5}.

Next we will extremize the action \eqref{s4d1cov} for the Poincare and global $AdS_5$ respectively.

\subsection{Poincare $AdS_5$ with Higher Derivatives}

For the time slice of Poincare $AdS_5$ we consider the form of the metric as given in (\ref{aidwjbnd}), and with that the action in \eqref{s4d1cov} becomes (upto some total derivatives)
\be \label{s4d1p}
\begin{split}
I_{4} = \int dz  \bigg[&\alpha_1  e^{4 \phi } + \alpha_2 e^{2\phi } (\partial_z\phi)^2 - \tilde b (\partial_z\phi)^4  -3b(\partial^2_z\phi)^2\bigg],
\end{split}
\ee
where we defined $\tilde b=3b+2a$ and also assumed that $\phi$ is a function of $z$ only. Extremizing the action in \eqref{s4d1p} we demand that the time slice of Poincare $AdS_5$ is a solution to that. In other words, $e^{\phi} = \ell / z $ extremizes the action if the following condition is satisfied
\be \label{solcond1p}
\alpha_1 \ell^4= \alpha_2 \ell^2 -6 a.
\ee

\subsection{Global $AdS_5$ with Higher Derivatives}

For time slice of global $AdS_5$ we again consider the metric as in (\ref{aidwjbndd}) and the corresponding action functional \eqref{s4d1cov}, turns out to be
\be \label{s4d1g}
\begin{split}
I_{4} = \int dr r^3 \bigg[\alpha_1  e^{4 \phi } + \alpha_2 e^{2\phi } \phi'^2 - \tilde b \phi'^4 +4 \tilde b \phi'^2 \phi'' -6b\phi''^2-3b  \phi'\phi''' \bigg]
\end{split}
\ee
where $\phi'= \partial_r\phi$ and we have also assumed that $\phi$ is a function of $r$ only.

It is straightforward to check that $e^{\phi} = 2\ell / (1-r^2)$ is a solution to the equation of motion for $\phi$ obtained by extremizing \eqref{s4d1g}, provided
\be \label{solcond1}
\alpha_1 \ell^4= \alpha_2 \ell^2 -6 a,
\ee
which is same as \eqref{solcond1p} and hence we prove that  the time slice of global $AdS_5$ is indeed obtained by extremizing  \eqref{s4d1g}.

\subsection{Excitation in Global $AdS_5$ with Higher Derivatives}

Consider excited states in CFT$_4$ dual to AdS$_5$ Schwarzschild black holes (\ref{AdSbhqwe}).
In Euclidean path integral analysis, we consider a spherically symmetry excitation and write its metric perturbation as
\be \label{pertdef5hd}
e^{\phi} = {2 \over (1-r^2)} \bigg( 1+ M ~ \beta(r) \bigg).
\ee
Working up to linear order in $M$, we substitute  \eqref{pertdef5hd} in the equation of motion for $\phi$ that follows from the action in \eqref{s4d1g} and solve for $\beta(r)$. We use the restriction on the parameters as in \eqref{solcond1p} for the zeroth order solution. Also demanding that the solution be regular at $r=1$ we check that
\be \label{etasol5hd}
\beta(r) = \eta(r)
\ee
is indeed a allowed solution, where $\eta(r)$ is given in (\ref{wwesqhjhj}).
Therefore we conclude that even in the presence of the higher derivative terms in \eqref{s4d1g}, once the condition \eqref{solcond1p} is maintained the first order perturbed metric of the AdS BH agrees with the extremization of $I_{4}$.

\section{Entanglement Entropy and Liouville Field}\label{EELiou}
In our approach with the Liouville action and the metric
\be
ds^2=e^{2\phi(z)}\left(dz^2+dx^2\right),
\ee
we compute entanglement entropy as a line integral along the geodesic $\gamma$ in the hyperbolic plane that is attached to the endpoints of the interval $l$
\be
S_l=\frac{c}{6}\int_\gamma e^{\phi(z)}ds \label{EEOAP}
\ee
and for a general geodesic parametrized by $(z(t),x(t))$, we have
\be
ds=\sqrt{x'^2+z'^2}dt.
\ee
Moreover, it is important to note that all our "optimized" vacuum solutions not only satisfy the Liouville equation but also
\be
\partial_x\phi(z)=\partial_z\phi(z)+e^{\phi(z)}=0.\label{Onshell}
\ee
Notice also, that because we are interested in the regularized curve, we can just compute the entanglement entropy by (twice) the integral from the boundary to some distance in the bulk (turning point of the geodesic). That implies, using \eqref{Onshell}
\be
S_l\simeq \frac{c}{3}\int^{\tilde{L}}_\epsilon e^{\phi(z)}dz=-\frac{c}{3}\int^{\tilde{L}}_\epsilon \partial_z\phi(z)dz=-\frac{c}{3}\left[\phi(z)\right]^{\tilde{L}}_\epsilon
\ee
This is clear for the vacuum solution
\be
\phi_0(z)=-\log\left(z\right)
\ee
and for $\tilde{L}=l$ we obtain the usual result for the entropy.\\
In general we can consider an arbitrary conformal transformation of the Liouville field of the "vacuum" by chiral and anti chiral functions $(w,\bar{w})\to (f(w),g(\bar{w}))$. Under such transformation, Liouville field itself transforms as
\be
\phi(f,g)=\phi(w,\bar{w})-\frac{1}{2}\log\left(f'(w)g'(\bar{w})\right).
\ee
This is still a solution of the Liouville equation with negative curvature (hyperbolic) and, in our approach, leads to a particular CFT state. Interestingly, we can then compute the entanglement entropy for such solution and after the line integral \eqref{EEOAP}, we obtain
\be
S_{l}=\f{c}{12}\log \left(\f{(f(w_1)-f(w_2))^2}{f'(w_1)f'(w_2)\epsilon^2}\right)+\f{\bar{c}}{12}\log\left(\f{(g(\bar{w}_1)-g(\bar{w}_2))^2}{g'(\bar{w}_1)g'(\bar{w}_2)\epsilon^2}\right).\label{EEGen}
\ee
Curiously, from the general solution of the Liouville equation, we can now see that that this result itself can also be written as a Liouville field and satisfies the Liouville equation but with positive curvature \cite{BHHM} and the space described by the end-points of the interval. It appears that these two Liouville fields can obtained form each other by simple analytic continuation (see also \cite{CzechC}) but the physical significance of this fact is far from obvious and remains to be elucidated.\\
Nevertheless, given \eqref{EEGen}, we can still apply the first law and compute the stress-tensor. Namely, if we set $w_2=w_1+l$ and $\bar{w}_2=\bar{w}_1+l$, we can expand for small interval $l$
 \be
 S_{l}=\f{c+\bar{c}}{6}\log\f{l}{\epsilon}-\f{l^2}{6}\left(\f{c}{12}\{f(w_1),w_1\}+\f{\bar{c}}{12}\{g(\bar{w}_1),\bar{w}_1\}\right)+O(l^3)
 \ee
 where the expressions in the brackets are the Schwarzian derivatives
 \be
 \{f(w),w\}=\f{f'''(w)}{f'(w)}-\f{3}{2}\left(\f{f''(w)}{f'(w)}\right)^2.
 \ee
On the other hand, we would like to extract this date from the original Liouville field (hyperbolic) that enters in the optimization procedure. This can be done as follows: Note that the entropy in the new geometry is computed by
\be
 S_{l}=\f{c}{6}\int_{\gamma}e^{\phi(w,\bar{w})}e^{-\frac{1}{2}\log\left(f'(w)g'(\bar{w})\right)}ds.
 \ee
 If we then consider the exponent of the change in the Liouville field, the stress tensor (Schwarzian derivative) can be read of from the simple equation
 \be
 \partial^2_we^{-\frac{1}{2}\log\left(f'(w)g'(\bar{w})\right)}=-\f{1}{2}\{f(w),w\}e^{-\frac{1}{2}\log\left(f'(w)g'(\bar{w})\right)},
 \ee
 and analogously for $g$.


\end{document}